\documentclass[prd,superscriptaddress,twocolumn,nopreprintnumbers,floatfix,longbibliography]{revtex4}

\usepackage{amssymb}
\usepackage{amsmath}
\usepackage{graphicx}
\usepackage{dcolumn}
\usepackage{hyperref}
\usepackage{color,units}
\usepackage{lineno}
\usepackage{xspace}
\usepackage{mathtools}
\usepackage{tensor}

\newcommand{\sur}{{\sc NRHybSur3dq8}\xspace}

\newcommand{\proposal}{proposal}
\newcommand{\target}{target}

\begin{document}

\title{Higher-order gravitational-wave modes with likelihood reweighting}

\author{Ethan Payne}
\email{ethan.payne@ligo.org}
\affiliation{School of Physics and Astronomy, Monash University, Clayton, VIC 3800, Australia}
\affiliation{OzGrav: The ARC Centre of Excellence for Gravitational-Wave Discovery, Clayton, VIC 3800, Australia}

\author{Colm Talbot}
\email{colm.talbot@monash.edu}
\affiliation{School of Physics and Astronomy, Monash University, Clayton, VIC 3800, Australia}
\affiliation{OzGrav: The ARC Centre of Excellence for Gravitational-Wave Discovery, Clayton, VIC 3800, Australia}

\author{Eric Thrane}
\email{eric.thrane@monash.edu}
\affiliation{School of Physics and Astronomy, Monash University, Clayton, VIC 3800, Australia}
\affiliation{OzGrav: The ARC Centre of Excellence for Gravitational-Wave Discovery, Clayton, VIC 3800, Australia}

\date{\today}

\begin{abstract}
The gravitational waveform of a merging stellar-mass binary is described at leading order by a quadrupolar mode.
However, the complete waveform includes higher-order modes, which encode valuable information not accessible from the leading-order mode alone.
Despite this, the majority of astrophysical inferences so far obtained with observations of gravitational waves employ only the leading order mode because calculations with higher-order modes are often computationally challenging.
We show how to efficiently incorporate higher-order modes into astrophysical inference calculations with a technique known as importance sampling.
There are two steps.
First, we carry out Bayesian parameter estimation using a computationally cheap leading-order-mode waveform, which provides an initial estimate of binary parameters.
Second, we weight the initial estimate using higher-order mode waveforms in order to fold in the extra information from the full waveform.
We use mock data to demonstrate the effectiveness of this method.
We apply the method to each binary black hole event in the first gravitational-wave transient catalog GWTC-1 to obtain posterior distributions and Bayesian evidence with higher-order modes.
Performing Bayesian model selection on the events in GWTC-1, we find only a weak preference for waveforms with higher-order modes, indicating that the events observed to date are not sufficiently loud to resolve the higher-order modes.
We discuss how this method can be generalized to a variety of other applications.
\end{abstract}

\maketitle

\section{Introduction}
The precise morphology of a gravitational waveform encodes a wealth of information about the binary that produced it.
Merging stellar-mass binaries are typically characterized by fifteen parameters: two mass parameters, six spin parameters, and seven extrinsic parameters, which describe the location and orientation of the binary with respect to the detector~\footnote{Binary neutron stars can be characterized by additional tidal parameters and all compact binaries can be characterized with additional eccentricity parameters, depending on their formation scenario.}.
Extracting binary parameters from gravitational-wave measurements enables tremendous science including sky maps for electromagnetic follow-up~\cite{GW170817_properties,GW170817_mma}, measurement of the neutron star equation of state~\cite{GW170817_tidal}, measurement of cosmological parameters~\cite{Schutz,GW170817_Hubble}, probing the fate of massive stars~\cite{mass_uc,mass,o2pop}, understanding the formation mechanisms of compact binaries~\cite{salvo,Stevenson,spin,GerosaBerti,FarrNature,Wysocki18,eccentricity}, and testing general relativity~\cite{GW150914_gr,GRB170817A}.

The parameters of compact binaries are estimated using Bayesian inference software~\cite{lalinference,PyCBCInference,bilby,RapidPE1,RapidPE2}.
The software employs nested sampling~\cite{Skilling2004}, Markov Chain Monte Carlo~\cite{Metropolis1953,Hastings1970,Hogg}, or adaptive mesh refinement~\cite{RapidPE1,RapidPE2} in order to construct posterior distributions for binary parameters and/or to calculate the Bayesian evidence.
Bayesian inference calculations in gravitational-wave astronomy are computationally demanding, and so significant research has been carried out in order to bring down the wall time of calculations, thereby enabling new science; see, e.g.,~\cite{smith,purrer,canizares,gpu_inference,gpu_rit}.

The computational demands of inference have created a premium for fast approximate gravitational waveforms or ``approximants''~\cite{IMRPhenomPv2,SEOBNRv2}.
Fast approximants have enabled breakthrough science.
However, the speed can come at a cost.
The current approximants most commonly used in gravitational-wave inference are constructed using only the leading order, $\ell=2$ modes in the spin-weighted spherical harmonic decomposition, although see~\cite{IMRPhenomHM}.
While these leading-order approximants provide reasonably good estimates of binary parameters, they do not incorporate all of the information in a gravitational waveform, and therefore provide an incomplete picture.
Inference with higher-order modes can provide tighter constraints than those obtained with leading-order waveforms alone.
In particular, higher-order modes are useful breaking degeneracy between binary parameters.
For example, the $\ell=|m|=2$ waveform is totally invariant under a transformation in which the polarization angle and phase of coalescence advance by $\pi/2$.
The ability to break this degeneracy is key to detecting gravitational-wave memory~\cite{memory}.
This is just one example highlighting the scientific potential of inference with higher-order modes.

Astrophysical inference with higher-order mode waveforms was first demonstrated in~\cite{Kumar,RapidPE2}, which applied a numerical relativity surrogate model~\cite{NRSur7dq2} to produce posterior distributions for GW150914, GW170104, GW170608, and GW170814.
More recently, adaptive mesh methods have been employed in order to derive posterior distributions and Bayesian evidence for GW170729 and other events in GWTC-1~\cite{gwtc-1} using a variety of higher-order-mode waveforms~\cite{Chatziioannou}.

In this paper we demonstrate a fast and effective method to calculate posterior distributions and Bayesian evidence for gravitational-wave signals with higher-order-modes using a technique known as ``importance sampling''; see, e.g.,~\cite{Robert&Casella,Liu}.
First, we carry out Bayesian parameter estimation using a low-cost, $\ell=|m|=2$ waveform, which yields an approximate answer on which we can improve.
In the second step, we calculate a weight factor for each posterior sample, which incorporates information from the more expensive higher-order-mode waveform.
Using the weights, we obtain the posterior and evidence, which we would have obtained if we had carried out the entire calculation using the more expensive higher-order mode waveform.

The remainder of this paper is organized as follows.
In Section~\ref{method}, we describe the reweighting formalism.
Then, in Section~\ref{results}, we show results obtained for both simulated data and for events in the LIGO/Virgo catalog, GWTC-1~\cite{gwtc-1,losc}.
Posterior samples and Bayesian evidence for every event in GWTC-1 are available on a companion web page~\cite{hom-git}, along with the code used in our analysis.
We provide a table summarizing the evidence of higher-order modes in GWTC-1.
We show that, while higher-order modes produce tighter constraints than $\ell=2$ analyses, there is not yet a strong signature of higher-order modes in published LIGO/Virgo detections. 
In Section~\ref{conclusions}, we discuss other possible directions for future research including novel applications of this reweighting technique, which may be useful for a variety of problems in astrophysics.
Finally, we include an appendix, Section~\ref{catalog_posteriors}, with the rest of the results from GWTC-1.

\section{Methodology}\label{method}
We show how to carry out Bayesian inference using an approximate ``\proposal'' likelihood ${\cal L}_{\O}(d|\theta)$ to obtain initial posterior samples, which are then reweighted using a more computationally expensive ``\target'' likelihood ${\cal L}(d|\theta)$.
Here, $d$ is the data and $\theta$ represents the parameters.
The \proposal\ likelihood is an approximation for the \target\ likelihood.
In order for reweighting to be efficient, the \proposal\ likelihood should be similar to the \target\ likelihood, so that the two likelihoods overlap significantly.
For demonstration purposes, we use as our \proposal\ waveform {\sc IMRPhenomD}~\cite{IMRPhenomD}, an aligned-spin, $\ell=|m|=2$ approximant, which is widely used in astrophysical inference thanks to its reliability and speed.
For our \target\ waveform, we use \sur~\cite{NRHybSur3dq8}, a numerical relativity surrogate model, which includes higher-order modes up to $\ell=4$ excepting $(4,\pm1)$ and $(4,0)$, but including $(5,\pm5)$, aligned spin, and mass ratios $m_2/m_1\geq 0.125$.

Our goal is to derive expressions for the \target\ posterior
\begin{align}
    p(\theta|d) = \frac{{\cal L}(d|\theta) \pi(\theta)}{\cal Z} ,
\end{align}
and the \target\ Bayesian evidence
\begin{align}
    {\cal Z} = \int d\theta {\cal L}(d|\theta) \pi(\theta) ,
\end{align}
written in terms of a fast-to-calculate, \proposal\ likelihood.

The \proposal\ quantities are linked to the \target\ quantities by a weight factor.
Multiplying the \target\ posterior by unity, we obtain
\begin{align}
    p(\theta|d) = & \frac{{\cal L}_{\O}(d|\theta)}{{\cal L}_{\O}(d|\theta)}
    \frac{{\cal L}(d|\theta) \pi(\theta)}{\cal Z} \nonumber\\
    = & w(d|\theta) \frac{{\cal L}_{\O}(d|\theta) \pi(\theta)}{\cal Z} .
\end{align}
Here,
\begin{align}
    w(d|\theta) \equiv \frac{{\cal L}(d|\theta)}{{\cal L}_{\O}(d|\theta)} ,
\end{align}
is the weight function.
Multiplying by unity again, we obtain the following expression for the evidence
\begin{align}
    {\cal Z} = & {\cal Z}_{\O} \int d\theta \, p_{\O}(\theta|d)
    \left(\frac{{\cal L}(d|\theta)}{{\cal L}_{\O}(d|\theta)}\right) \nonumber\\
    = & \frac{{\cal Z}_{\O}}{n}\sum_k^n w(d|\theta_k) .
\end{align}
The second line replaces the integral with a discrete sum over $n$ \proposal\ posterior samples; see~\cite{intro}.

Carrying out Bayesian inference with the \proposal\ likelihood, we obtain ``\proposal\ posterior samples'' for the distribution
\begin{align}
    p_{\O}(\theta|d) = \frac{{\cal L}_{\O}(d|\theta) \pi(\theta)}{{\cal Z}_{\O}} ,
\end{align}
where ${\cal Z_{\O}}$ is the \proposal\ evidence.
We generate our \proposal\ samples using the {\sc Bilby}~\cite{bilby} implementation of {\sc CPNest}~\cite{cpnest}.
We compute the weights using likelihoods marginalized over the coalescence time and reference phase (see, e.g.,~\cite{intro}) in order to avoid differences in the definition of these parameters between waveform models.

In a previous iteration of this work~\cite{arxiv}, we employed a slightly different approach.
Instead of marginalizing over time and phase, we chose the values of these parameters that maximized the overlap ${\cal O}$ of the proposal waveform and target waveform: 
\begin{align}
    {\cal O} \equiv \max_{t,\phi} \frac{\langle h_{\O}^+, h^+ \rangle + \langle h_{\O}^\times, h^\times \rangle}{\sqrt{\left(\langle h_{\O}^+,h_{\O}^+\rangle + \langle h_{\O}^\times,h_{\O}^\times\rangle\right) \left(\langle h^+,h^+\rangle + \langle h^\times,h^\times\rangle\right)}} .
\end{align}
Here, $h_{\O}^{+,\times}$ are the plus and cross components of the \proposal\ waveform while $h^{+,\times}$ are the plus and cross components of the \target\ waveform.
The angled brackets denote noise-weighted inner products.
This approach, used subsequently in other publications, e.g.,~\cite{isobel}, produces qualitatively similar results to marginalizing over ($t$,$\phi$), albeit with reduced computational efficiency.
Moreover, the marginalization approach is better justified from a statistical point of view.

Weighting each sample by $w(d|\theta)$, and renormalizing, we convert the \proposal\ posterior samples into \target\ posterior samples.
The application of weight factors has the effect of reducing the effective number of samples~\cite{Kish,Elvira}
\begin{align}
    n_\text{eff} = \frac{\left(\sum_k w_k\right)^2}{\sum_k w_k^2} .
\end{align}
After reweighting, it is therefore prudent to calculate $n_\text{eff}$ in order to determine that there are a suitably large number of samples.
It is straightforward to generate more weighted posterior samples by simply combining the results from multiple \proposal\ analyses run in parallel.

The method of likelihood reweighting outlined here is similar to the procedure of ``recycling'' commonly used to study the population properties of compact objects; see, e.g.,~\cite{o2pop,intro}.
Previous applications of recycling have, in effect, carried out reweighting to change the {\em prior} in post-processing.
The principle here is the same, except we change the likelihood.
Our formalism can be straightforwardly extended to simultaneously alter the prior (informed by astrophysics) and likelihood (using more sophisticated waveforms).
Reweighting is an application of importance sampling.
For additional details, readers are referred to~\cite{Robert&Casella,Liu}.

\section{Results}\label{results}
\begin{figure*}[t!]
    \centering
    \includegraphics[width=0.56\linewidth]{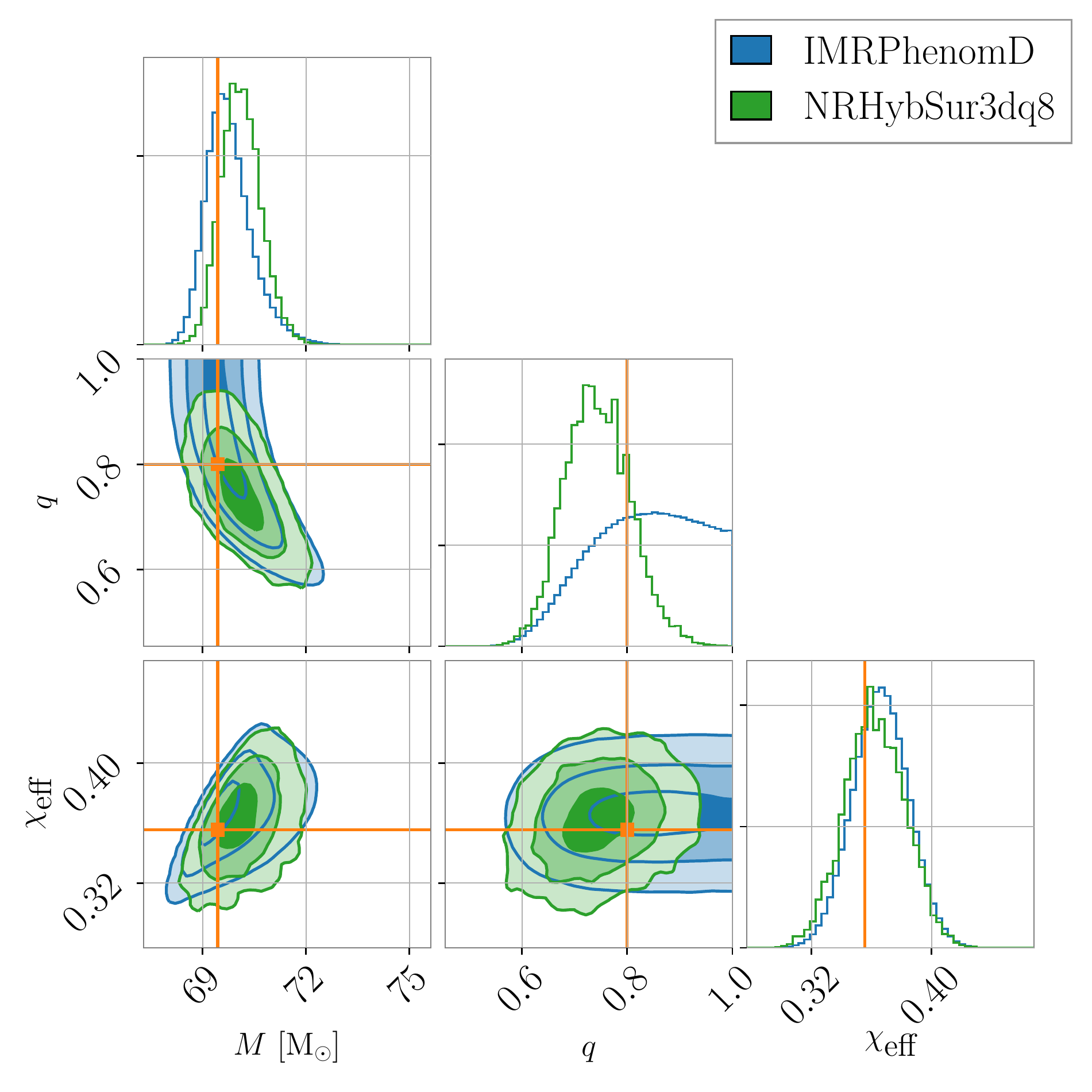}
    \includegraphics[width=0.405\linewidth]{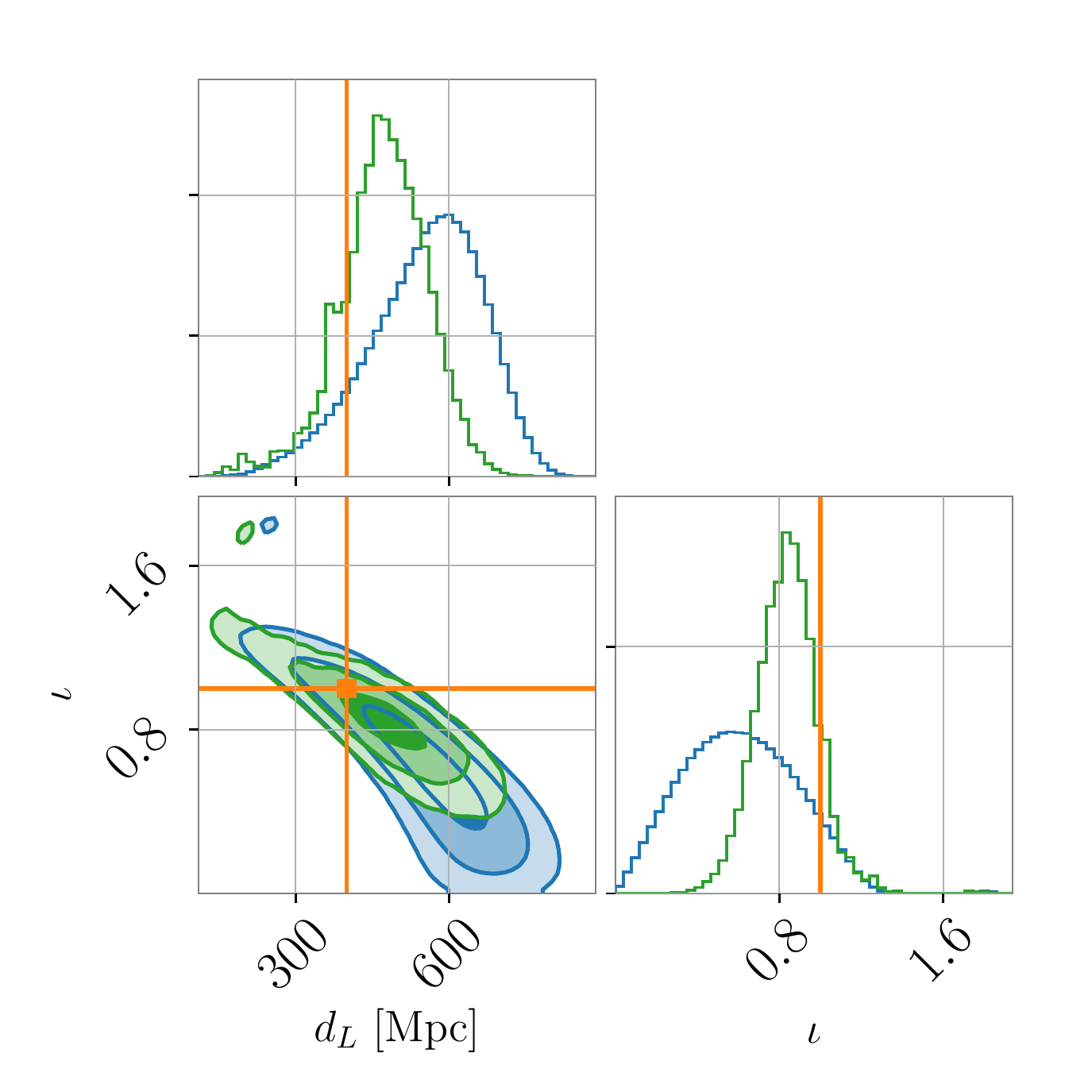}
    \caption{
    Posterior distributions for a simulated binary black hole waveform.
    The blue distributions show the posteriors obtained using the approximant {\sc IMRPhenomD}~\cite{IMRPhenomD}, which includes only the dominant $\ell=|m|=2$ modes.
    The green distribution shows the posteriors obtained using the approximant \sur~\cite{NRHybSur3dq8}, which includes higher-order modes.
    The true parameters are indicated with orange lines.
    Left: intrinsic binary parameters (total mass, mass ratio, and effective aligned spin). Right: extrinsic parameters (luminosity distance and binary inclination).
    }
    \label{fig:comparison_corner_plot}
\end{figure*}

We demonstrate likelihood reweighting  using a simulated binary black hole merger signal injected into Gaussian noise.
We assume a two-detector LIGO network operating at design sensitivity~\cite{aligo}.
Using \sur, we inject a binary black hole waveform.
The binary, located at a luminosity distance $d_L=\unit[400]{Mpc}$, has chirp mass $\mathcal{M}=30 M_\odot$ and mass ratio $q=m_2/m_1=0.8$.
The dimensionless aligned spins are $\chi_1=0.4,\chi_2=0.3$.
The signal has a network matched-filter signal-to-noise ratio of $\rho_\text{mf}=55$.
In Fig.~\ref{fig:comparison_corner_plot} we provide a corner plot showing the posterior distribution and credible intervals obtained for this simulated event.
The blue shades indicate the posterior derived using our \proposal\ {\sc IMRPhenomD} waveform while the green shades indicates the posterior obtained after reweighting with our \target\ \sur waveform.
The darkness of the contours indicate credible intervals at $1\sigma,2\sigma,3\sigma$.
The true values of each parameter are indicated by orange markers.

While both posteriors includes the true parameter values, the blue {\sc IMRPhenomD} posterior is broad in comparison to the green \sur posterior.
For some parameters, the posterior shrinks dramatically when we add information from higher-order modes.
Higher-order modes break the degeneracy between distance and inclination while improving our ability to measure the mass ratio.
This, in turn, improves our estimation of the spins.

We calculate the Bayes factor comparing the hypothesis that the data are best fit by \sur to the hypothesis that they are best fit by {\sc IMRPhenomD}.
This is a measure of degree to which the data prefer a model including higher-order modes \footnote{To be clear, we are not proposing $\ell|m|=22$ gravitational waves as a serious alternative to general relativity. Rather, we are using this Bayes factor in order to ascertain if we can resolve the higher-order mode content predicted by general relativity.}.
The log Bayes factor for our simulated event are reported in the first row of Tab.~\ref{tab:BF}.
We include also the ``efficiency,'' the number of effective samples used in each calculation, normalized by the number of \proposal\ samples.

\begin{center}
    \begin{table}
        \begin{tabular}{ |c|c|c| } 
            \hline
            event & $\ln\text{BF}$ & $n_\text{eff}/n$ \\\hline
            simulated & 9.01 & $2.2\times10^{-3}$\\\hline
            GW150914 & -0.20 & 0.31\\
            GW151012 & 0.40 & 0.36 \\
            GW151226 & 0.01 &  0.82\\
            GW170104 & 0.01 & 0.57 \\
            GW170608 & -0.16 & 0.73\\
            GW170729 & 1.23 & 0.13\\
            GW170809 & -0.09 & 0.67\\
            GW170814 & 0.19 & 0.48\\
            GW170818 & 0.31 & 0.74 \\
            GW170823 & -0.21 & 0.69 \\\hline
            GWTC-1 & 1.49 & N/A\\\hline
        \end{tabular}
        \caption{
            The log Bayes factor and the ``efficiency,''  equal to the number of effective samples $n_\text{eff}$ divided by the initial number of samples $n$.
        }
        \label{tab:BF}
    \end{table}
\end{center}

For the simulated event, the signal-to-noise ratio is sufficiently large to ``detect'' the presence of higher-order modes with a high significance $\ln\text{BF}=9.01$.
Since this simulated event has a high signal-to-noise ratio, the ratio of the effective number of samples to the number of \proposal\ samples is small, $\approx0.2\%$.
Thus, a large number of \proposal\ runs ($\approx1000$) are required in order to produce a well-sampled \target\ posterior.
Fortunately, these \proposal\ runs are embarrassingly parallel, which means that the wall time is no longer than a single \proposal\ inference run, provided sufficient computational resources are available.
It should also be noted that we are able to get posterior samples and evidence for all of the events in GWTC-1 with just a few parallel runs.
We further note that it may be possible to improve efficiency by restricting the \proposal\ prior based on early returns from the \target\ posterior.
The events in GWTC-1 have lower signal-to-noise ratio, and so the reweighting procedure is much more efficient.

\begin{figure*}[t!]
    \centering
    \includegraphics[width=0.56\linewidth]{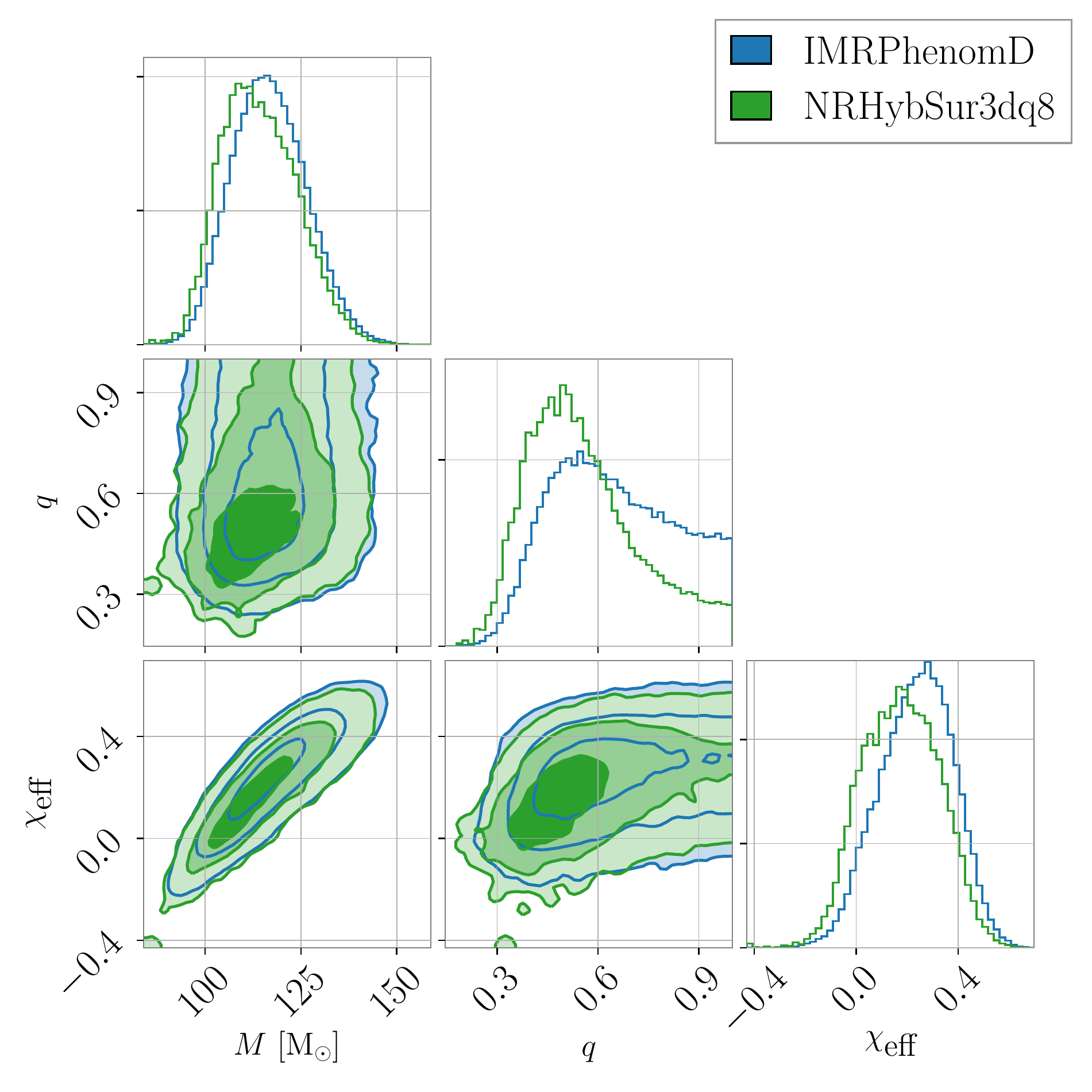}
    \includegraphics[width=0.405\linewidth]{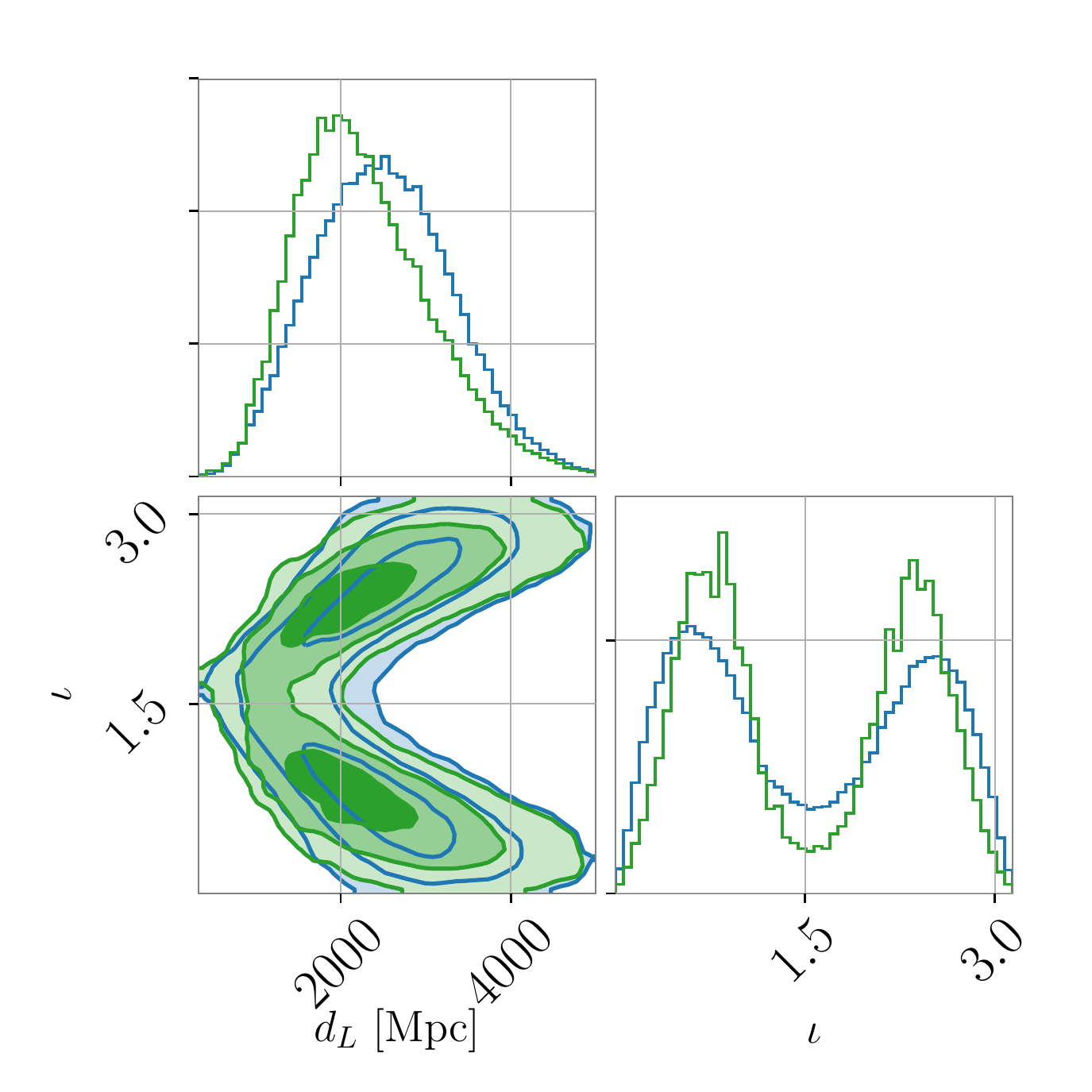}
    \caption{
    Posterior distributions for GW170729.
    The blue distributions show the posteriors obtained using the $\ell=|m|=2$ approximant {\sc IMRPhenomD}~\cite{IMRPhenomD}.
    The green distribution shows the posteriors obtained using the approximant \sur~\cite{NRHybSur3dq8}, which includes higher-order modes.
    Left: intrinsic binary parameters (total mass, mass ratio, and effective aligned spin). Right: the extrinsic parameters, luminosity distance and  inclination angle.
    }
    \label{fig:GW170729}
\end{figure*}

We now turn toward real data in the GWTC-1 catalog.
For each event, we produce: a set of posterior samples with weights, the \proposal\ Bayesian evidence (obtained with {\sc IMRPhenomD}), the \target\ Bayesian evidence (obtained with \sur), the \sur/{\sc IMRPhenomD} Bayes factor, and corner plots with credible intervals.
The full results are available here:~\cite{hom-git}.
Key summary statistics are provided in Tab.~\ref{tab:BF}.
For illustrative purposes, we also include in Fig.~\ref{fig:GW170729} the corner plot for GW170729, the event with the greatest support for higher-order modes ($\ln\text{BF}=1.23$).
Our \target\ posterior for GW170729 is qualitatively similar to previous results from~\cite{Chatziioannou}.
We observe increased support for non-unity mass ratio $q=m_2/m_1$, slightly more support for zero-spin, and changes to the posterior distributions of the extrinsic parameters.
An additional validation of the results for GW170729 are included in Appendix~\ref{validation}.

\section{Discussion}\label{conclusions}
As the gravitational-wave catalog grows, higher-order modes will provide a more precise view of the properties of compact objects.
Higher-order modes become increasingly important for systems: with large mass ratios $q\lesssim0.5$, with significant spin, and with large total mass~\cite{vijay}.
The method described here will be particularly useful for exploring such events.
In order to make use of likelihood reweighting, it is necessary to employ a suitable \proposal\ model, capable of producing posterior samples, which provide a reasonable approximation of the true distribution.
For inference with higher-order modes, this appears to be achievable using quadrupolar approximants.

One must exercise a degree of caution when choosing the \proposal\ likelihood to make sure it is suitable for the problem at hand.
For example, binary mergers with very low mass ratios $q\lesssim 1/6$ (not yet observed by LIGO/Virgo) can produce so much higher-order mode content that the quadrupolar approximants yield biased parameter estimates~\cite{bustillo,varma}.
There are two telltale signs, which indicate that the \proposal\ likelihood is poorly suited to the problem.
First, the efficiency $n_\text{eff}/n$ will be low even for modest signal-to-noise ratio because the \proposal\ samples are all in the tail of the \target\ likelihood distribution.
The sample(s) closest to the \target\ likelihood peak will be weighted as far more important than the other samples.
Second, the reweighted posterior distributions will be peak at the edges of the sampled region of parameter space.
By comparing the \target\ posterior to the \proposal, it is straightforward to determine if the \target\ posterior is railing against the edge of the \proposal\ samples~\footnote{Of course, if the \target\ posterior is multi-modal, there is an additional risk that the \proposal\ posterior might miss a mode, which would not be apparent through visual inspection. We have not observed such multi-modal behavior in the examples considered in this manuscript, but one can guard against this failure mode through $pp$ tests as described in the appendix.}.
If these two tests indicate a poorly chosen \proposal\ likelihood, one should devise a better one.

We foresee a number of useful applications for likelihood reweighting including: inferences about gravitational-wave memory~\cite{memory}, noise models that add complexity beyond the usual Gaussian assumption, treatment of calibration errors, and inference with computationally expensive waveforms that include tidal effects or eccentricity~\footnote{After this paper appeared on the arxiv, our method was used in~\cite{isobel}.}.

\section{Acknowledgements}
We thank Rory Smith and Greg Ashton for sharing a beta version of their parallelized nested sampling algorithm, used for benchmarking in the appendix.
We thank Rory Smith for helping to debug our parallel inference runs.
We thank Katerina Chatziioannou, Richard O'Shaughnessy, and Vijay Varma for helpful comments.
We thank Moritz H{\"u}bner, and the {\sc Bilby} team for support.
This is document LIGO-P1900128.
EP, ET, and CT are supported by ARC CE170100004.
ET is supported by ARC FT150100281.
This research has made use of data, software and/or web tools obtained from the Gravitational Wave Open Science Center (https://www.gw-openscience.org), a service of LIGO Laboratory, the LIGO Scientific Collaboration and the Virgo Collaboration. LIGO is funded by the U.S. National Science Foundation. Virgo is funded by the French Centre National de Recherche Scientifique (CNRS), the Italian Istituto Nazionale della Fisica Nucleare (INFN) and the Dutch Nikhef, with contributions by Polish and Hungarian institutes.

\bibliography{refs}

\begin{appendix}

\section{Validation}\label{validation}
\begin{figure*}[t!]
    \centering
    \includegraphics[width=0.56\linewidth]{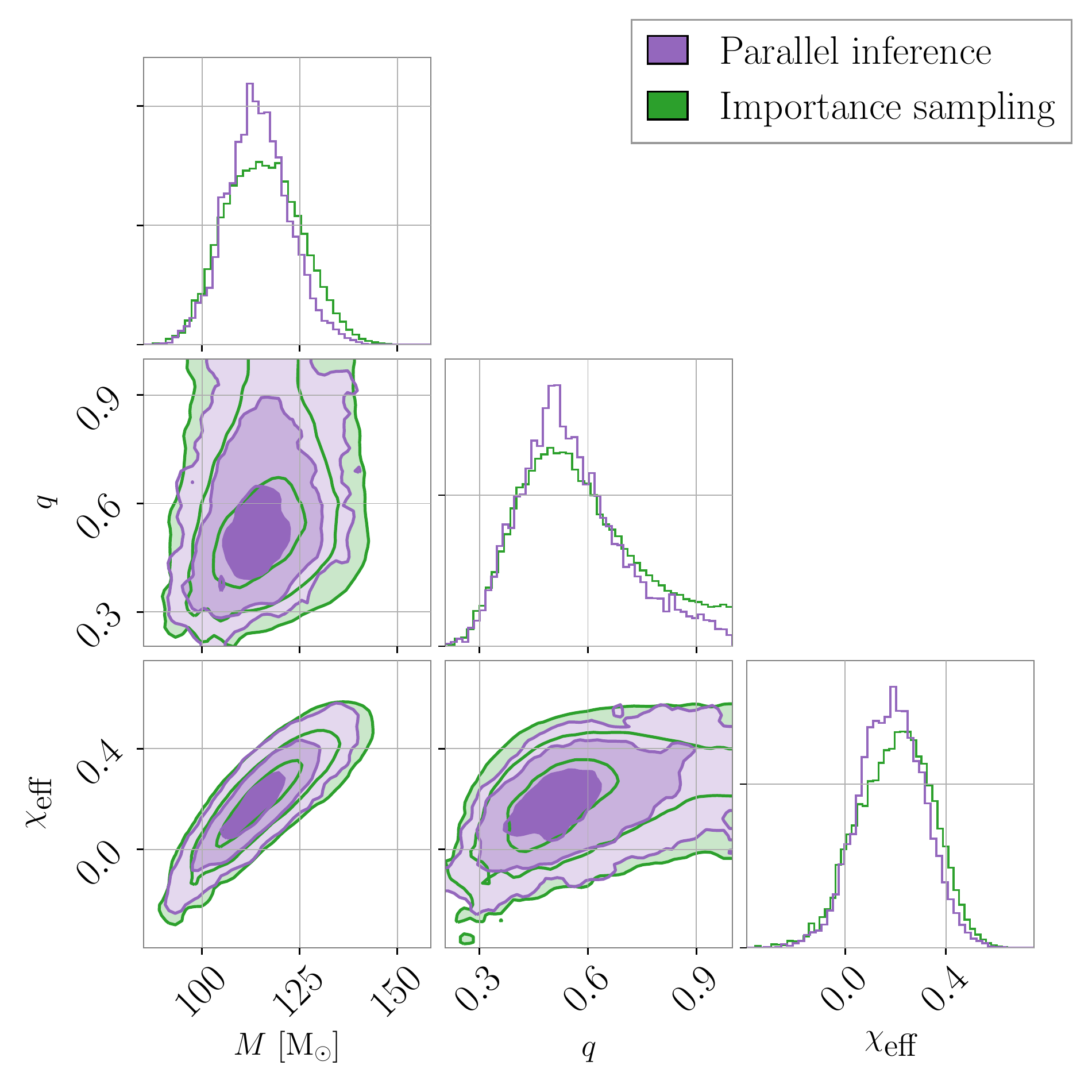}
    \includegraphics[width=0.405\linewidth]{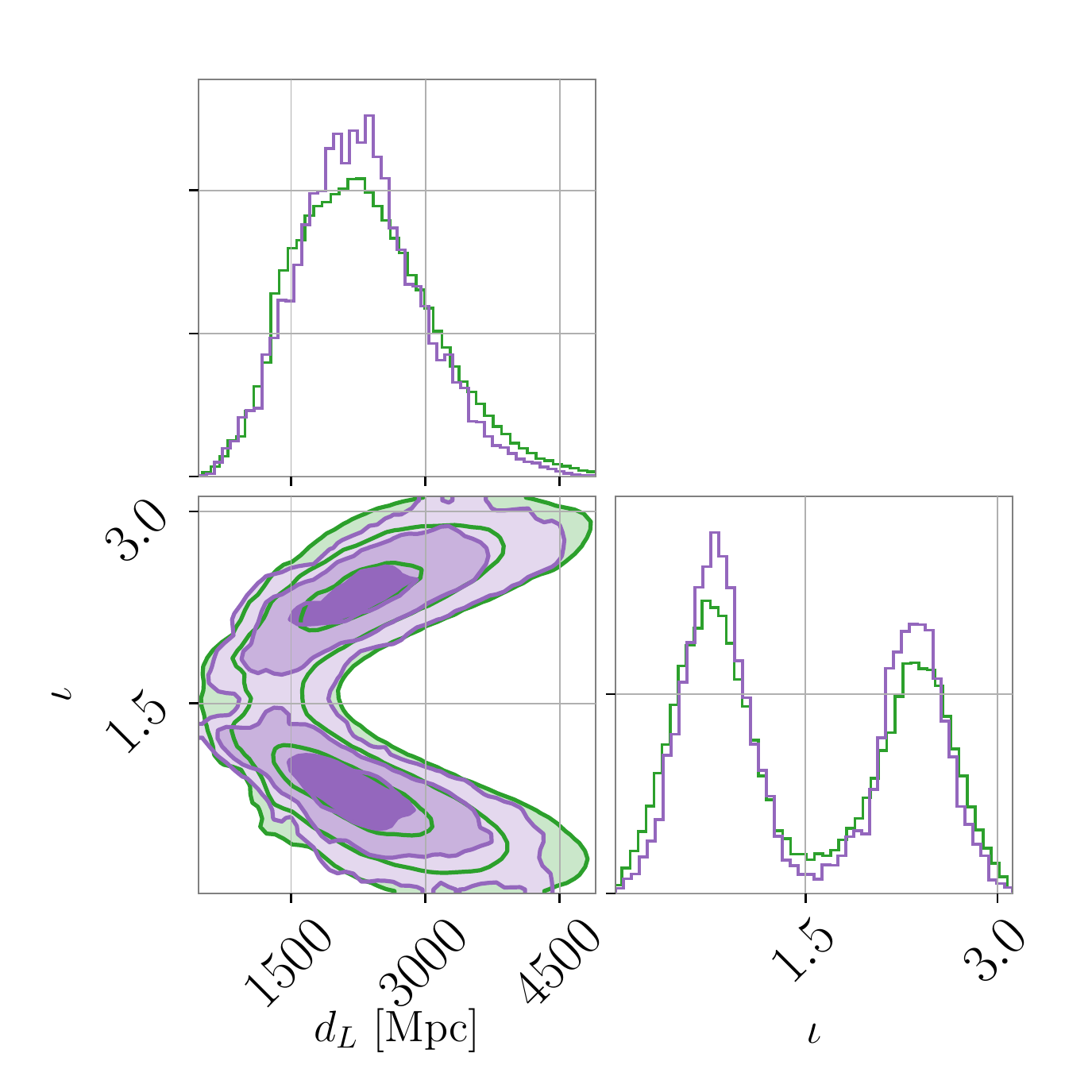}
    \caption{A comparison of the posterior distributions for GW170729 calculated using reweighting (green) and direct sampling (purple)~\cite{smith}.
    The two methods yield similar results.
    The remaining differences, which are typical for results obtained using two different implementations of nested sampling, are indicative of the systematic error from reweighting.}
    \label{fig:GW170729_compare}
\end{figure*}

In order to test the reweighting procedure, we calculate the posterior for GW170729 using direct sampling and compare the results obtained with reweighting.
The direct sampling method is computationally expensive.
However, we are able to make use of a recent work by Smith \& Ashton~\cite{SmithAshton}, who have created a parallelized implementation of the nested sampler, {\sc dynesty}~\cite{dynesty}.
The calculation ran for 25 hours on 830 cores, which suggests it would have taken more than two years to complete on a single core.
This is in contrast to our calculation, which took approximately 5 hours on 25 cores to generate the target samples and 9 hours on 450 cores to calculate weights (66 core days in total).
In Fig.~\ref{fig:GW170729_compare}, we plot the posterior for GW170729 obtained using reweighting (green) alongside the posterior obtained using the parallel inference code from~\cite{smith} (purple).
The results are in approximate agreement, with small differences that are typical of the differences that arise when a posterior is calculated with two different implementations of nested sampling, for example, {\sc dynesty} and {\sc CPNest}~\cite{cpnest}.
It is difficult to ascertain, which result is a more accurate representation of the true posterior distribution, but the level of disagreement should be taken as indicative of the  systematic error from reweighting.

\begin{figure}[t!]
\centering
    \includegraphics[width=1\linewidth]{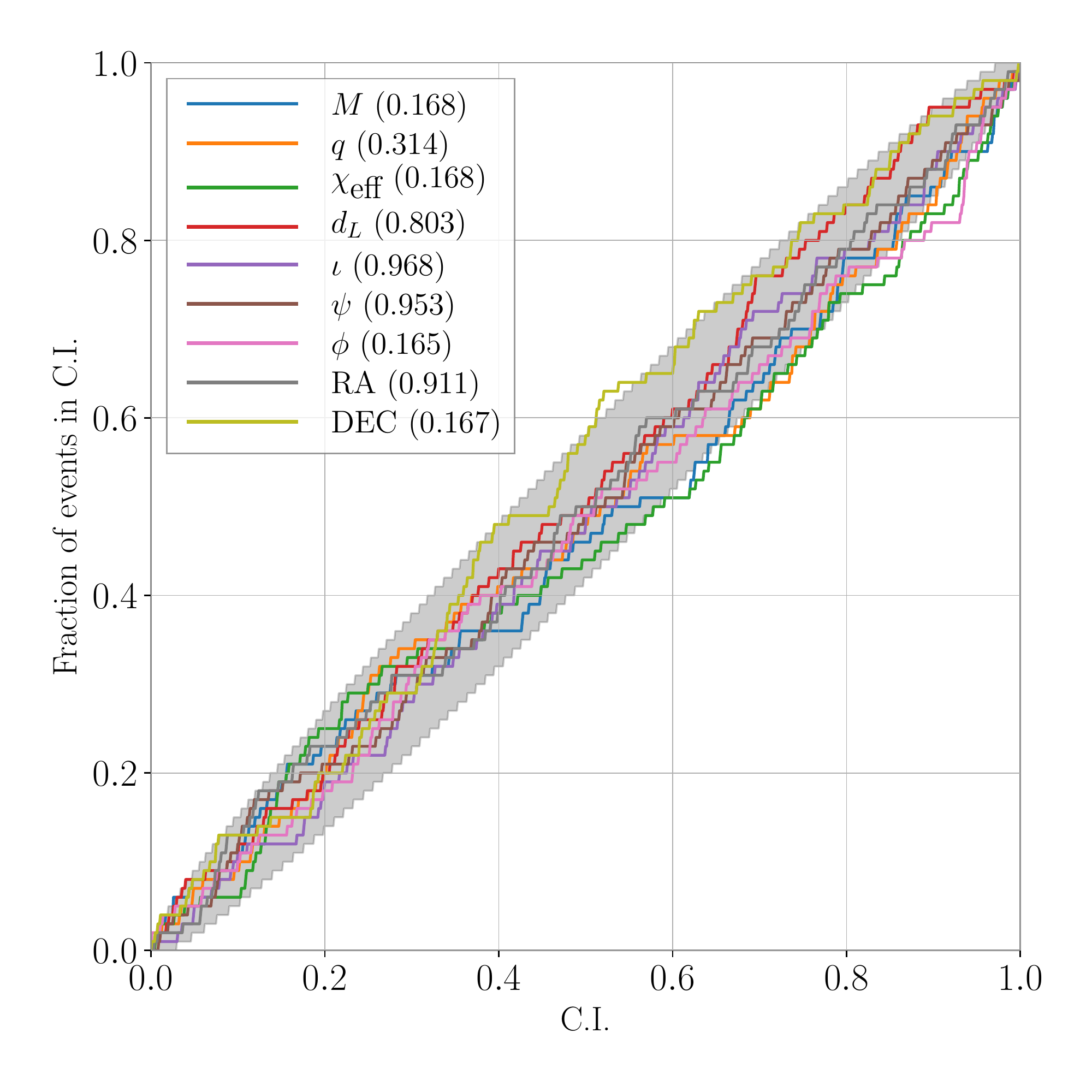}
    \caption{
    A ``$pp$ plot,'' showing the fraction of 100 simulated events (vertical axis) for which the true parameter value falls within a given confidence interval (horizontal axis).
    The gray region shows the typical fluctuations for a single parameter.}
    \label{fig:pp}
\end{figure}

In Fig.~\ref{fig:pp}, we include a ``$pp$ plot,'' which provides another means of validating the reweighting method.
The $pp$ plot shows the fraction of 100 simulated events (vertical axis), for which the true parameter value falls within a given confidence interval (horizontal axis).
Ideally, the colored traces (representing different astrophysical parameters) should be consistent with a slope=1 diagonal line.
The gray region indicates the typical fluctuations (90\% credible interval) for a single parameter due to the finite sample size.
The distribution of colored traces about the slope=1 line indicates that the reweighting method is producing reliable posteriors to within the precision we can test. 

\section{Results from GWTC-1}\label{catalog_posteriors}
In this appendix, we present posterior distributions and credible intervals for the rest of the events in GWTC-1.
In each figure, the blue distributions show the posteriors obtained using the $\ell=|m|=2$ approximant {\sc IMRPhenomD}~\cite{IMRPhenomD}.
The green distribution shows the posteriors obtained using the approximant \sur~\cite{NRHybSur3dq8}, which includes higher-order modes $\ell\leq4$  excepting $(4,\pm1)$ and $(4,0)$, but including $(5,\pm5)$.
The intrinsic binary parameters (total mass, mass ratio, and effective aligned spin) are on the left.
The extrinsic parameters, luminosity distance and  inclination angle, are on the right.

\begin{figure*}[t!]
    \centering
    \includegraphics[width=0.56\linewidth]{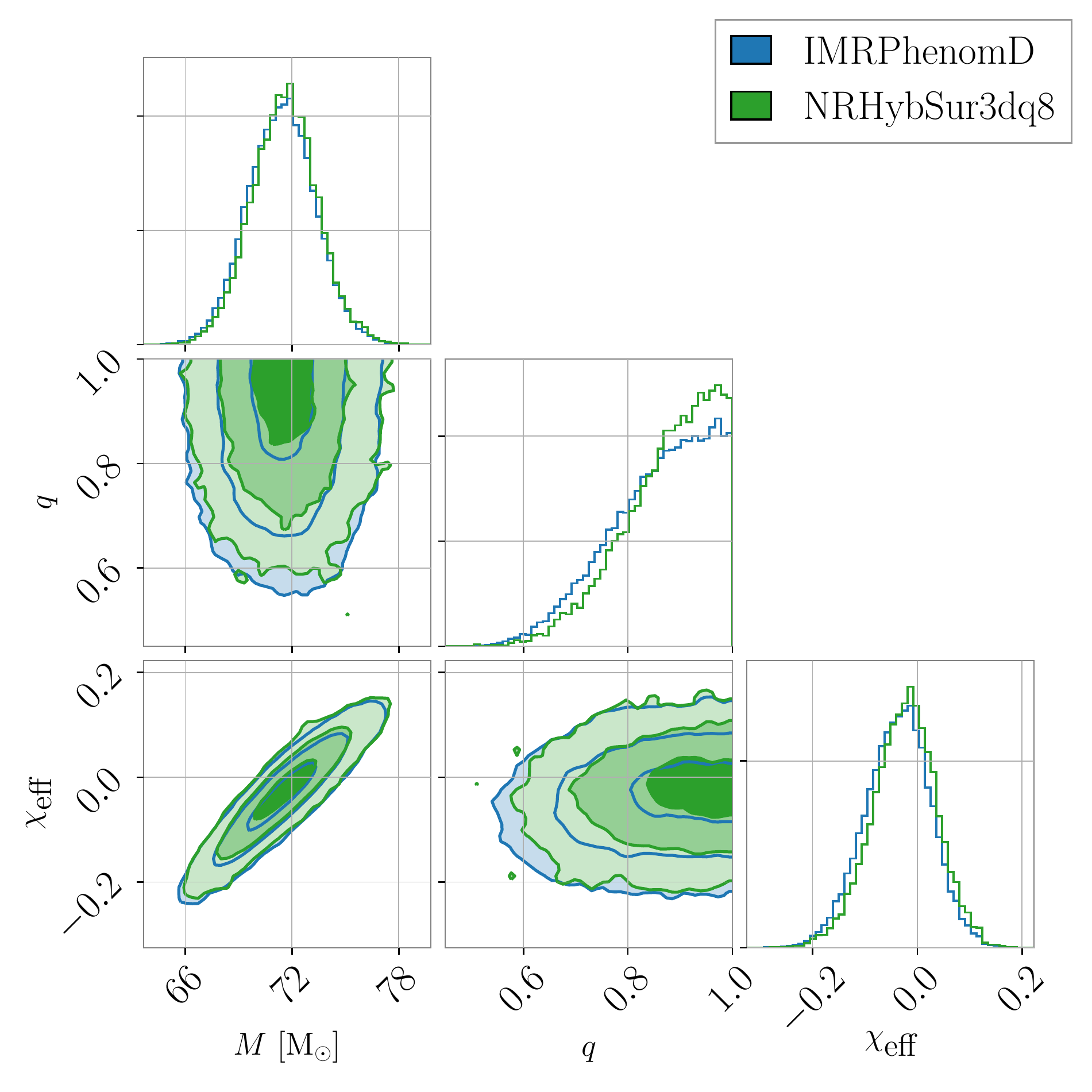}
    \includegraphics[width=0.405\linewidth]{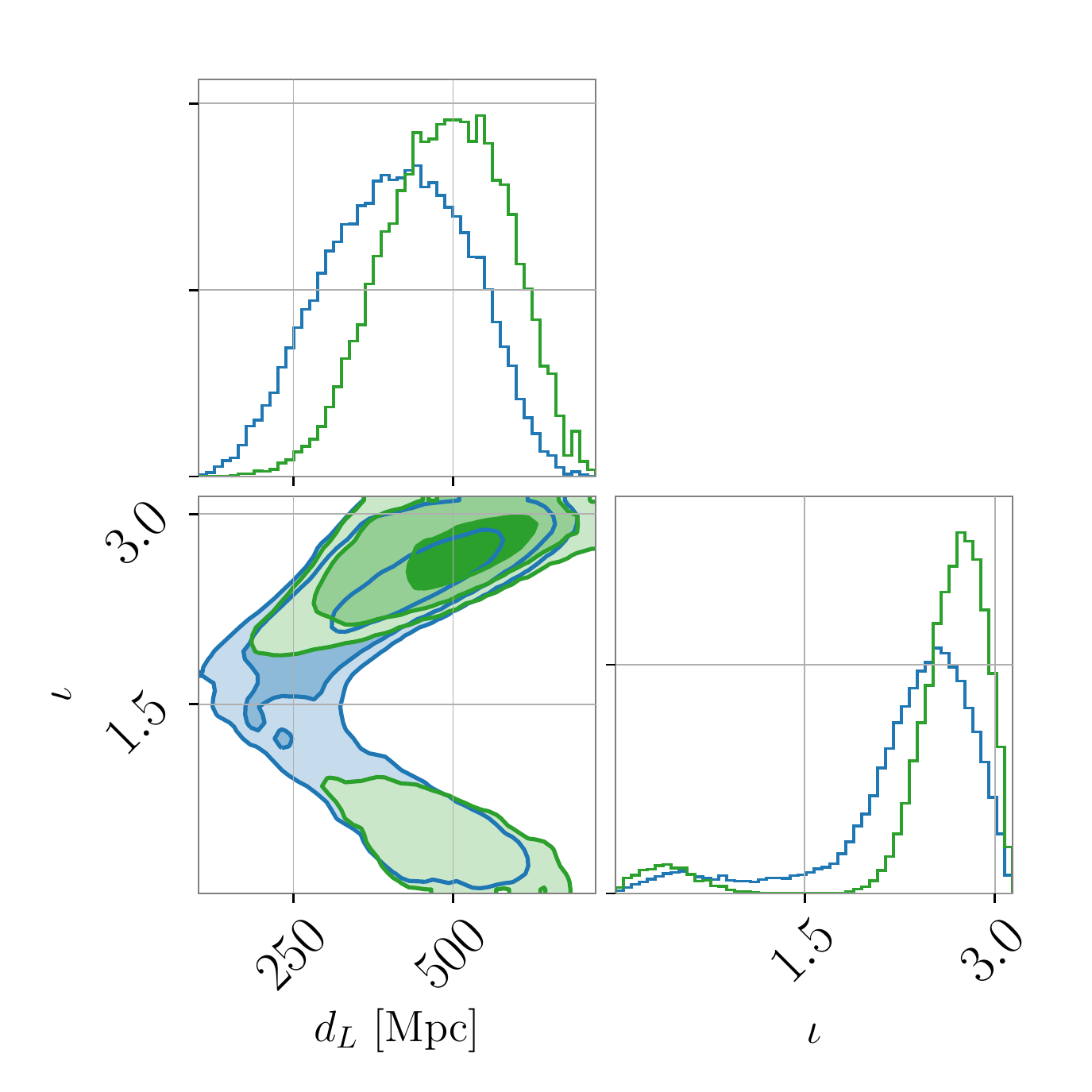}
    \caption{
    GW150914. This event is interesting because the negative $\ln\text{BF}$ indicates that higher-order modes are not preferred over $\ell=2$ waveforms.
    It is not surprising to obtain some negative log Bayes factors due to noise fluctuations since the higher-order mode signal is so small.
    The \sur posterior for inclination angle is pulled toward face-off where higher-order mode emission is reduced.
    The distance posterior shifts further away.
    }
    \label{fig:GW150914}
\end{figure*}

\begin{figure*}[t!]
    \centering
    \includegraphics[width=0.56\linewidth]{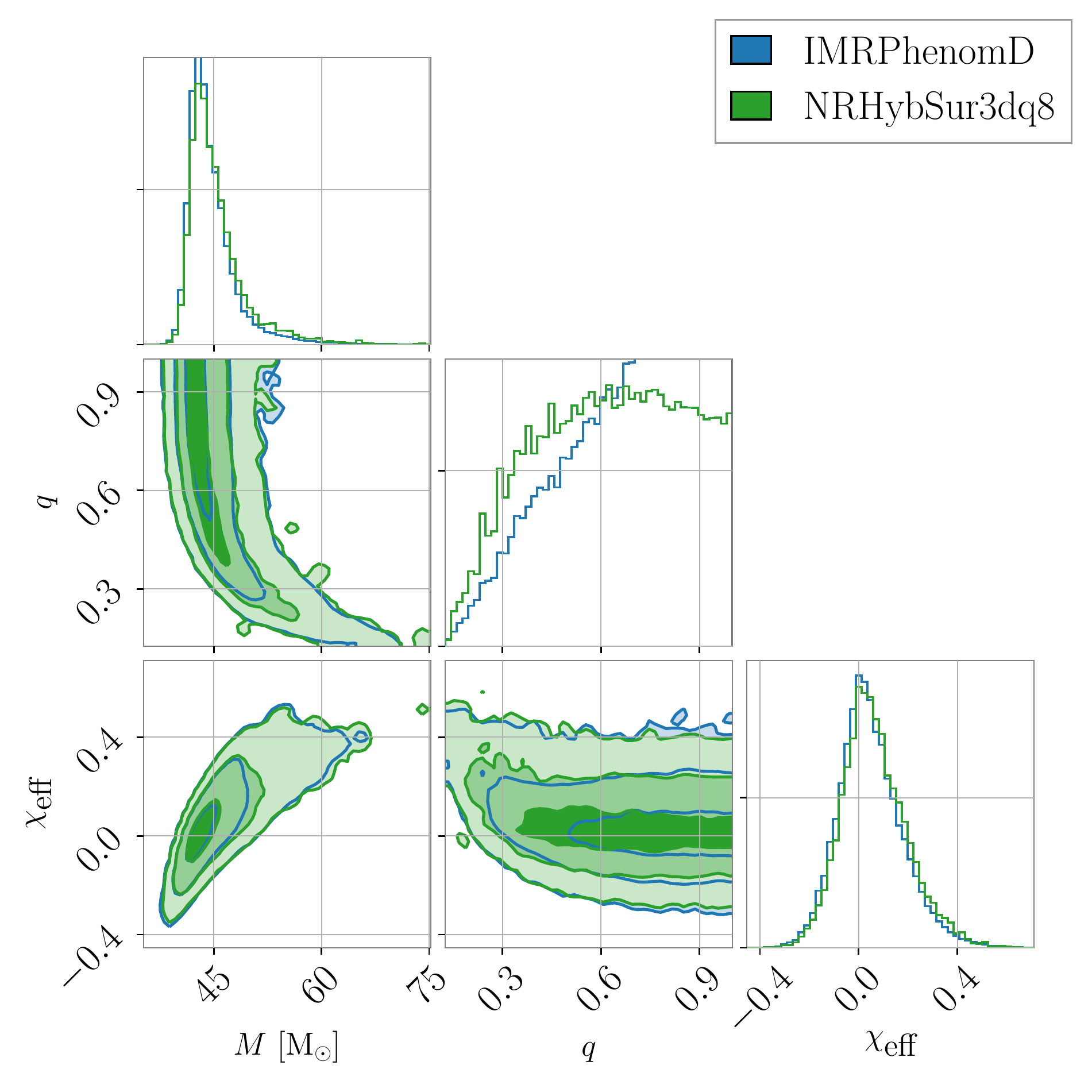}
    \includegraphics[width=0.405\linewidth]{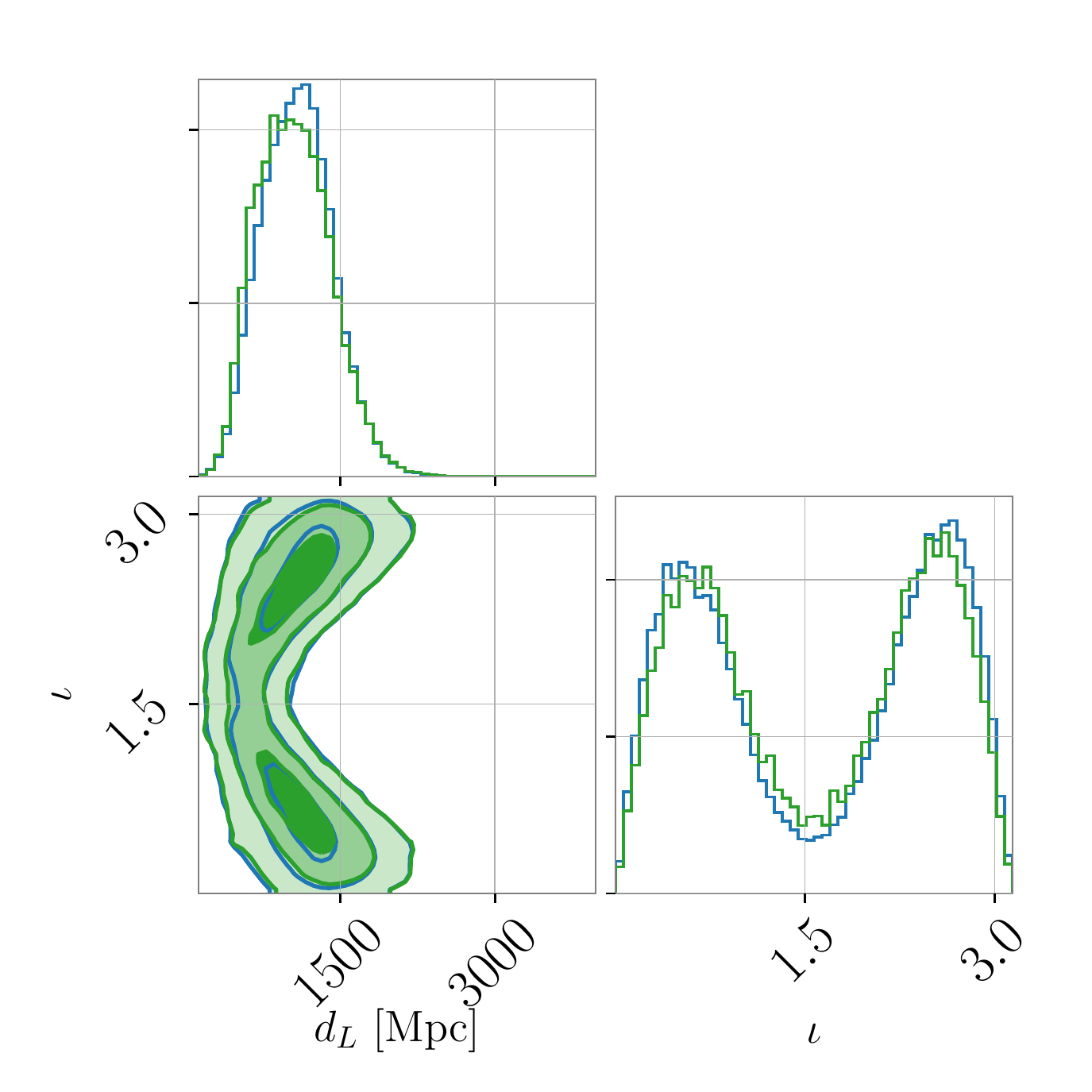}
    \caption{
    GW151012}
    \label{fig:GW151012}
\end{figure*}

\begin{figure*}[t!]
    \centering
    \includegraphics[width=0.56\linewidth]{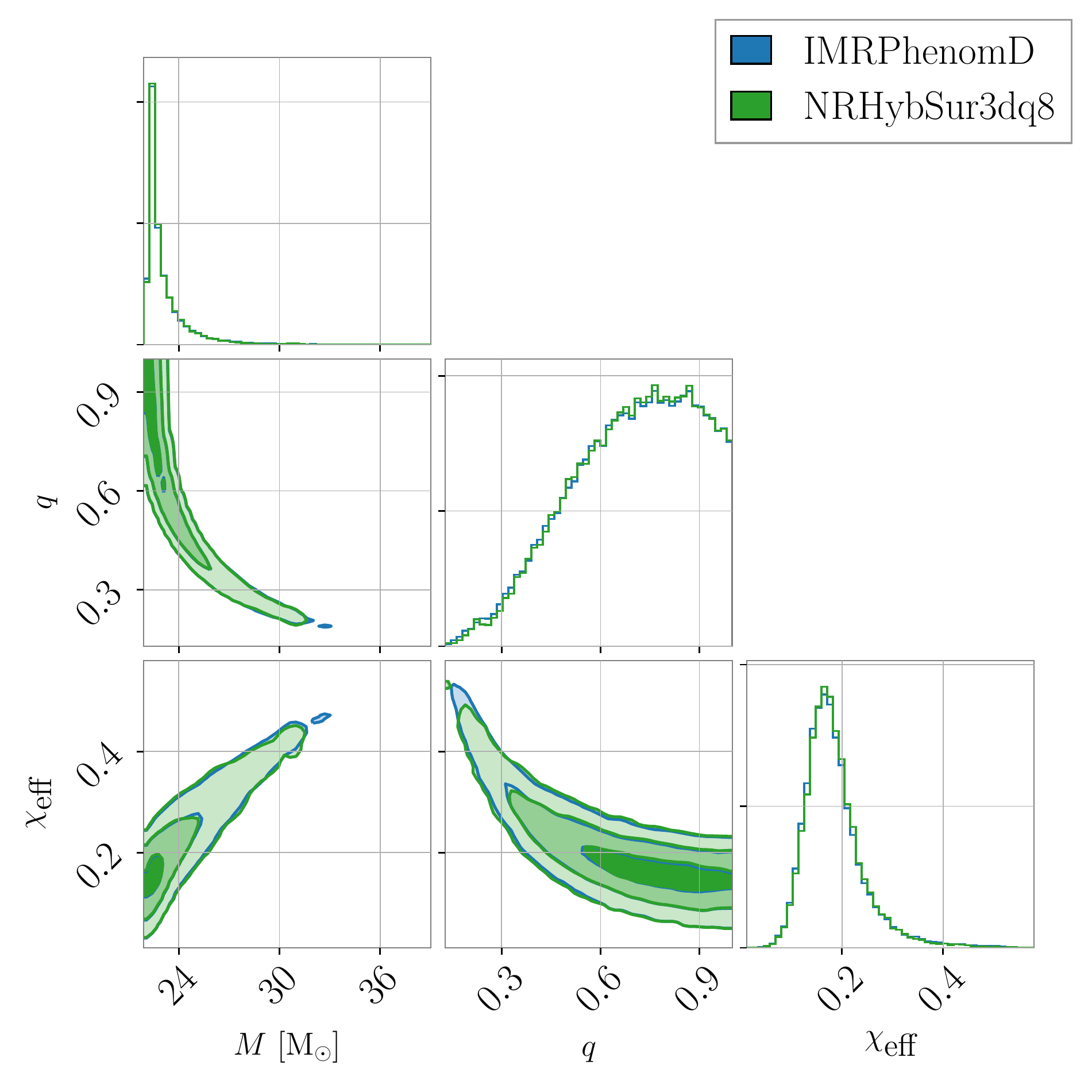}
    \includegraphics[width=0.405\linewidth]{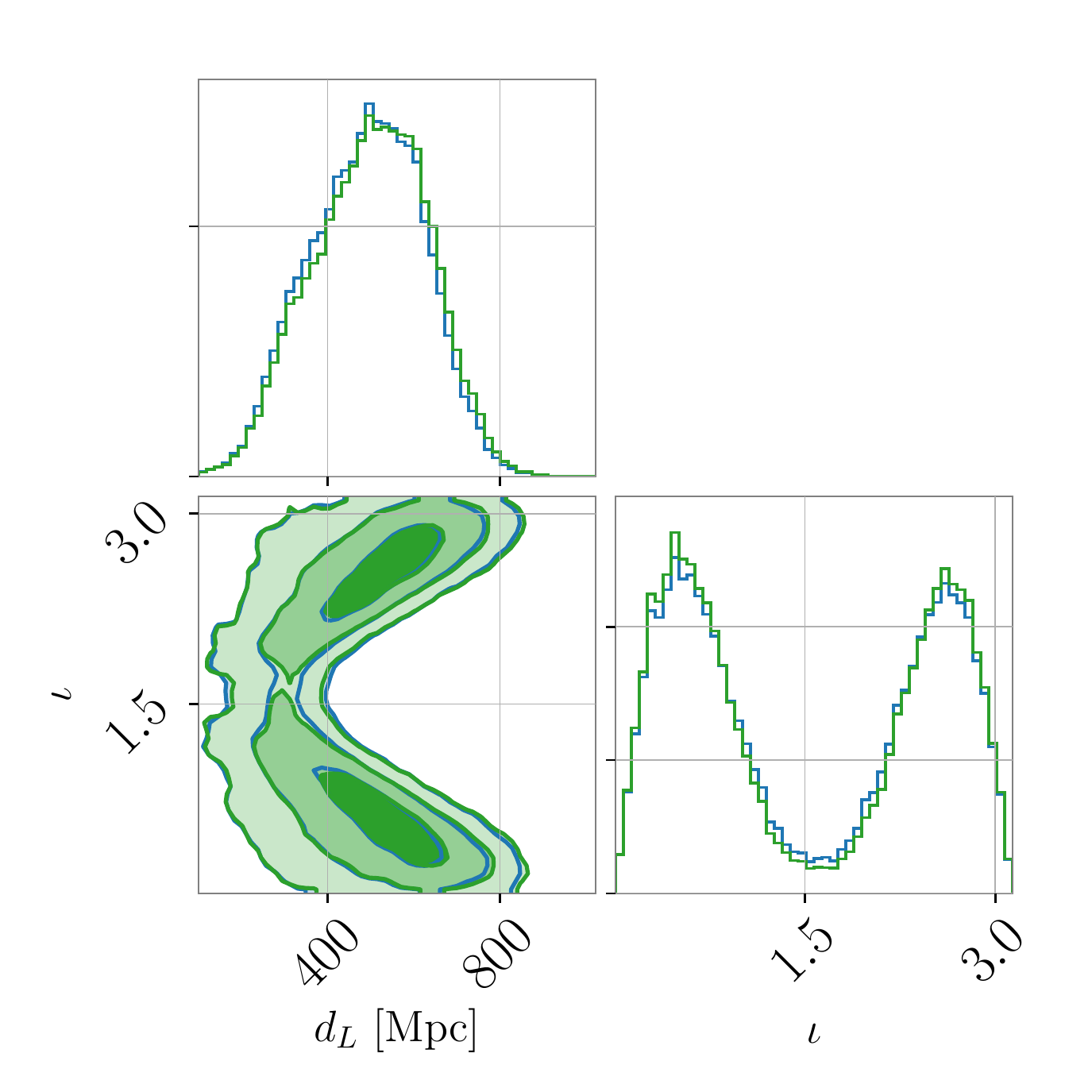}
    \caption{
    GW151226}
    \label{fig:GW151226}
\end{figure*}

\begin{figure*}[t!]
    \centering
    \includegraphics[width=0.56\linewidth]{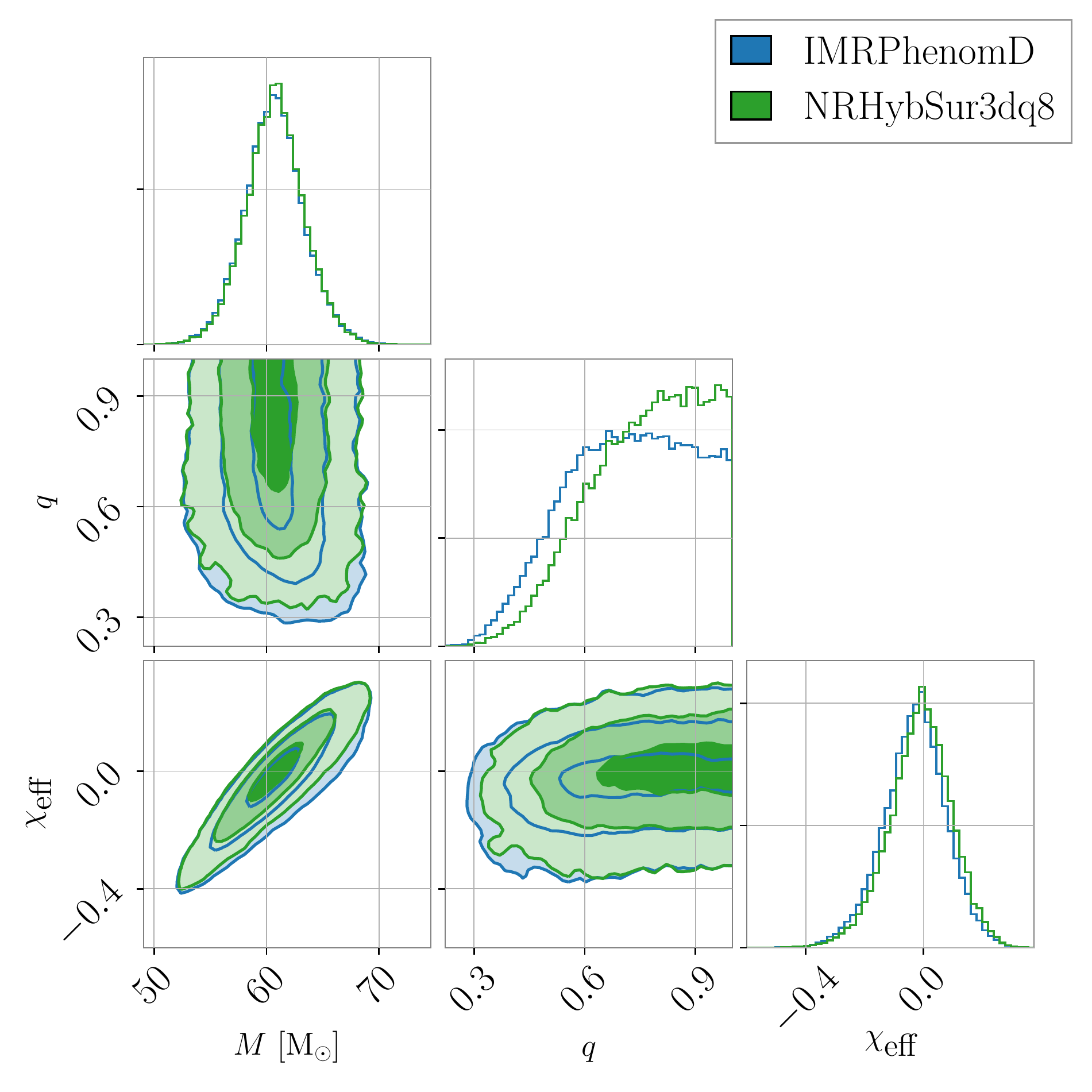}
    \includegraphics[width=0.405\linewidth]{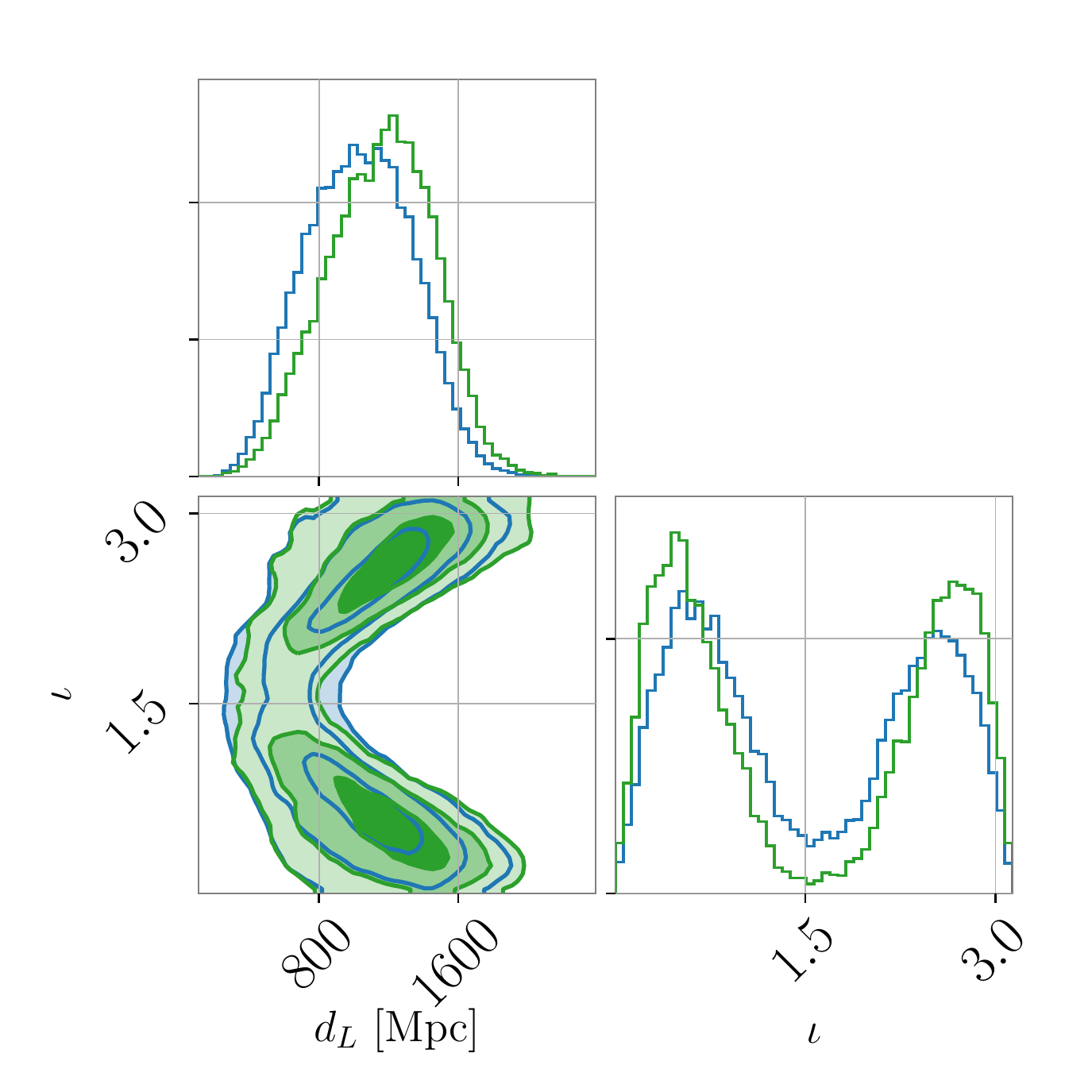}
    \caption{
    GW170104}
    \label{fig:GW170104}
\end{figure*}

\begin{figure*}[t!]
    \centering
    \includegraphics[width=0.56\linewidth]{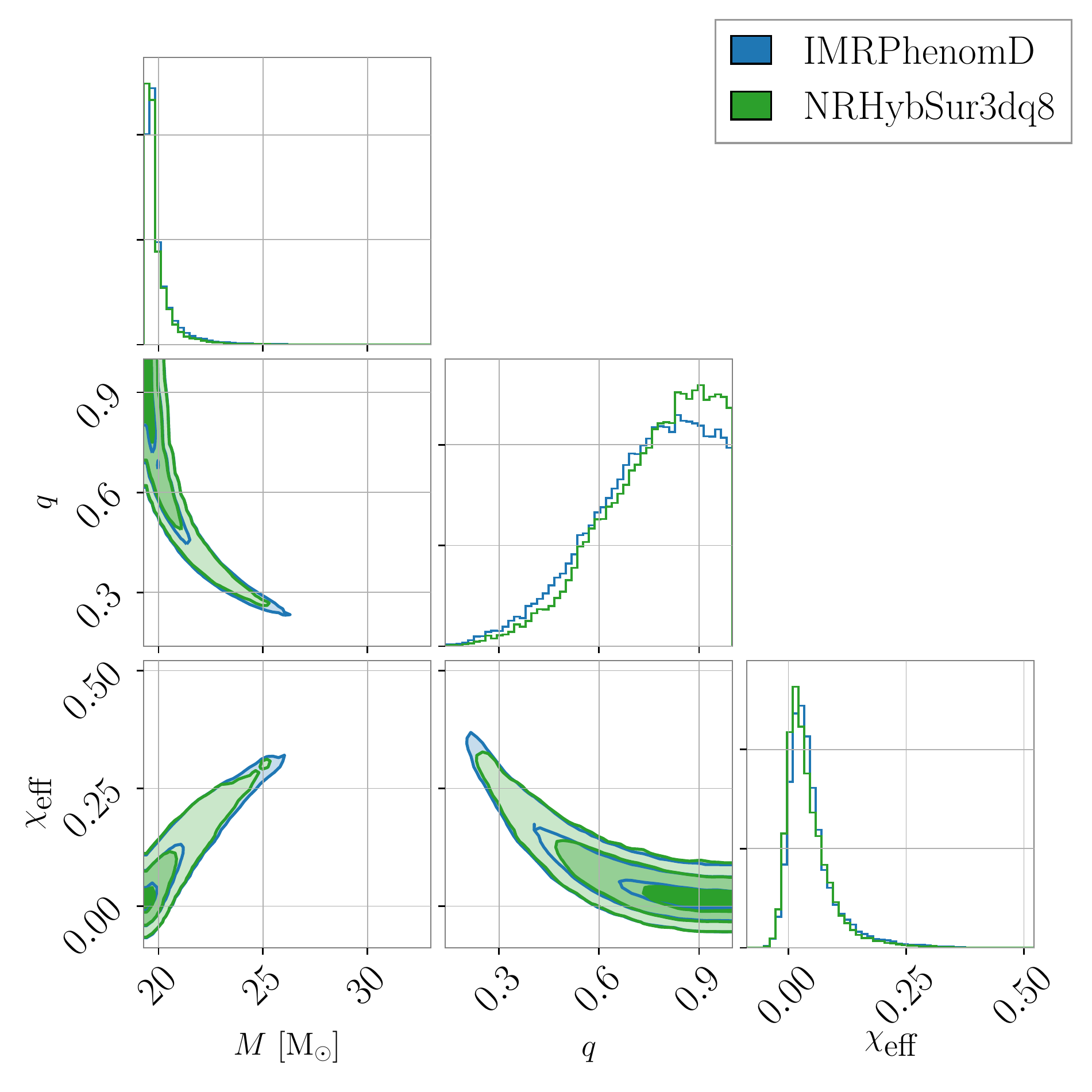}
    \includegraphics[width=0.405\linewidth]{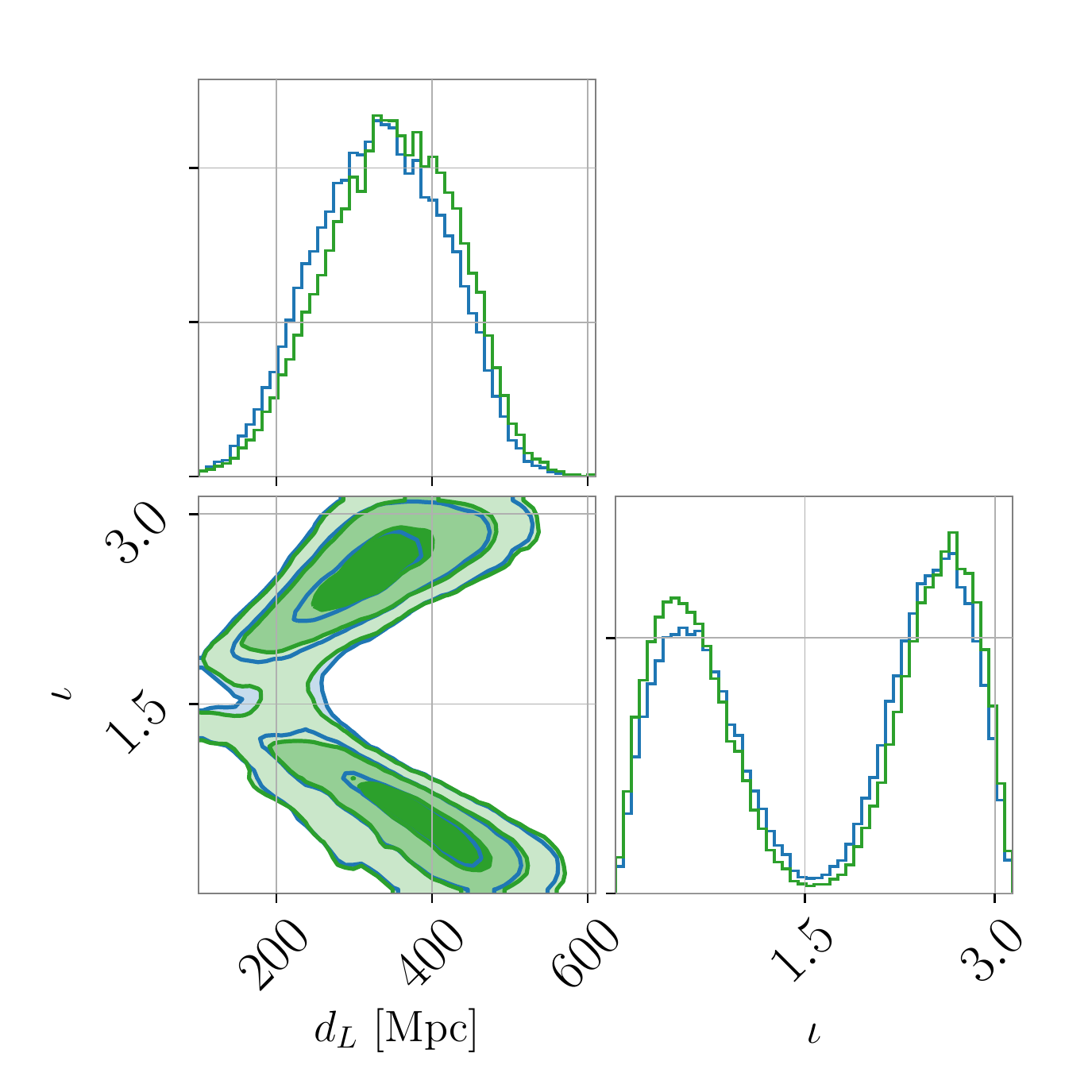}
    \caption{
    GW170608}
    \label{fig:GW170608}
\end{figure*}

\begin{figure*}[t!]
    \centering
    \includegraphics[width=0.56\linewidth]{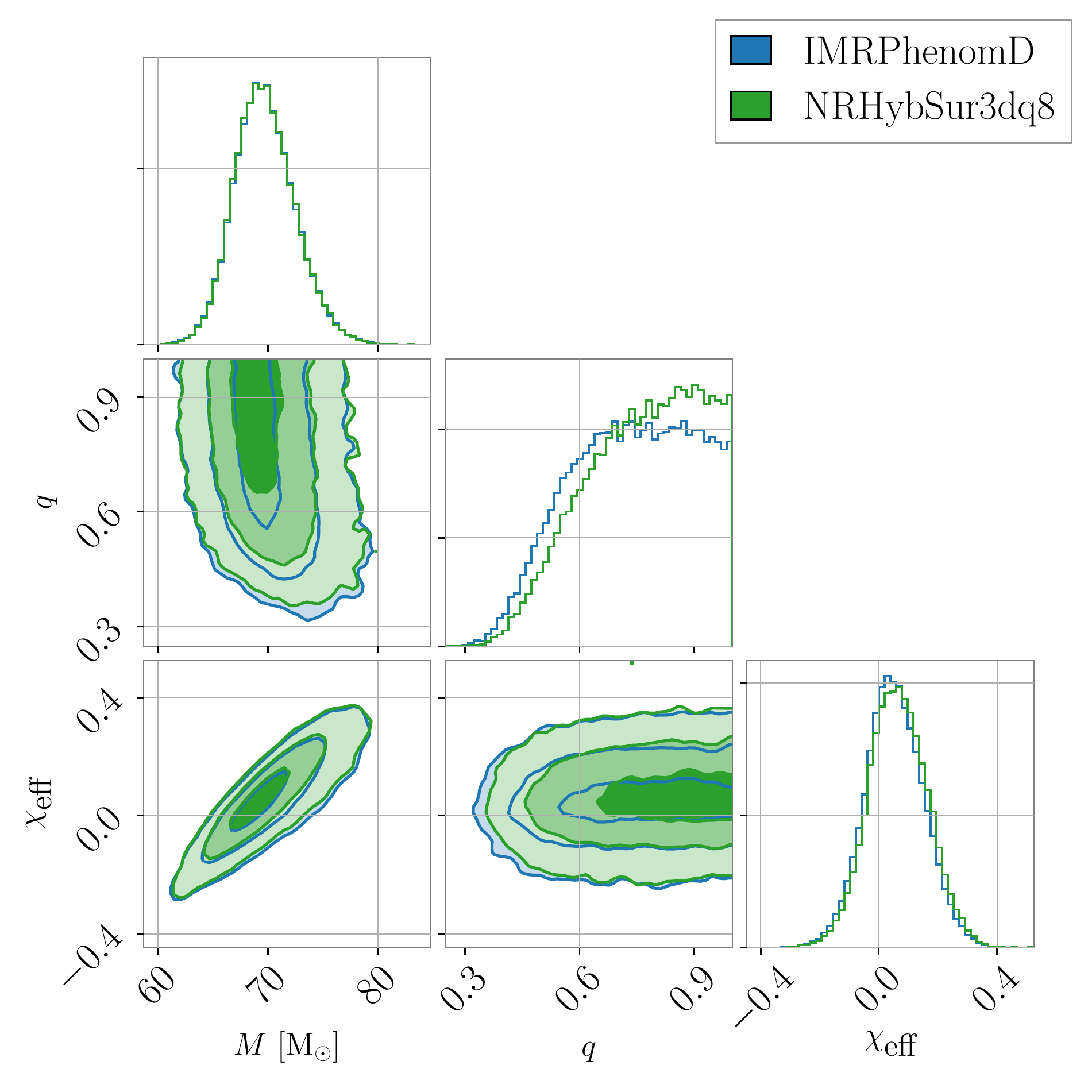}
    \includegraphics[width=0.405\linewidth]{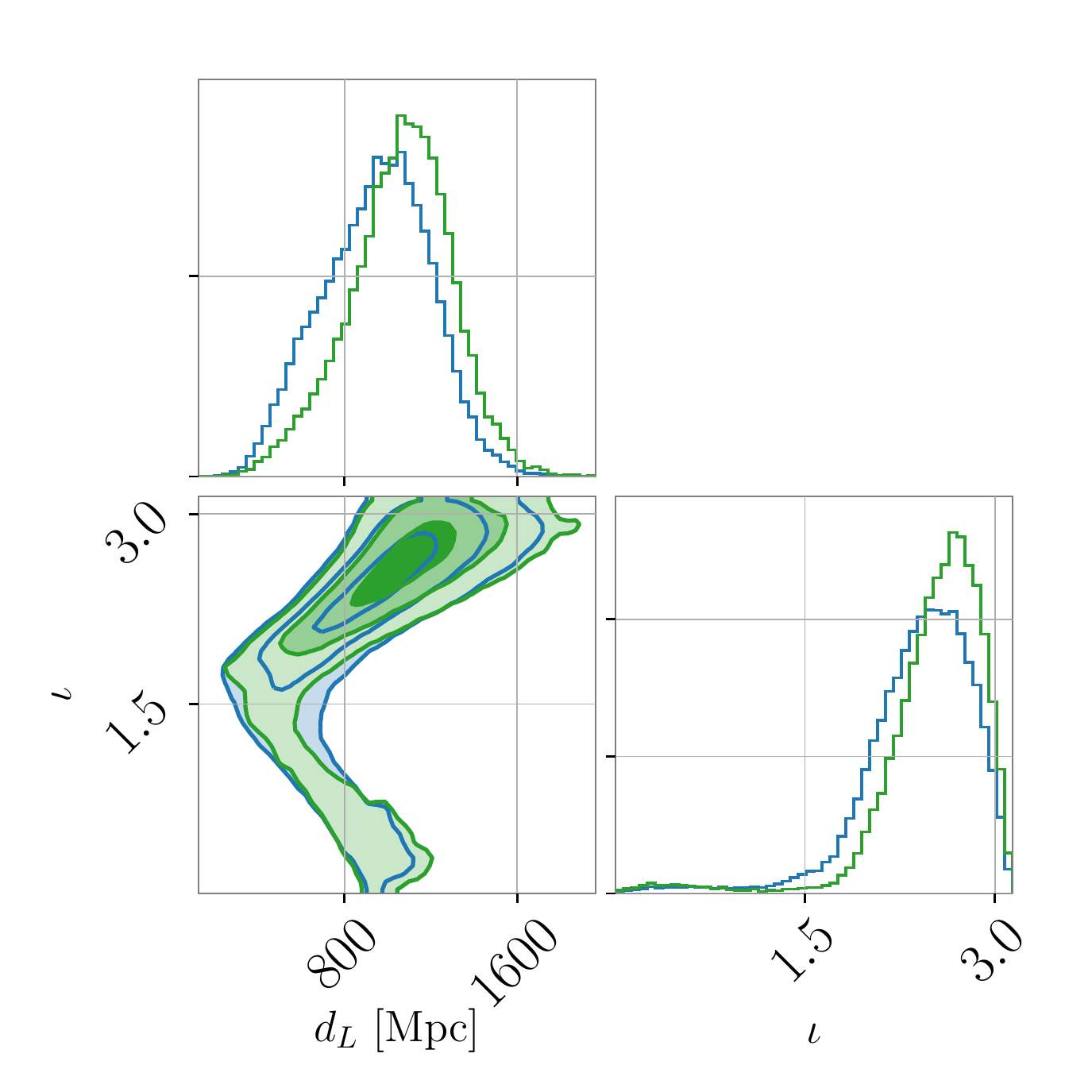}
    \caption{
    GW170809}
    \label{fig:GW170809}
\end{figure*}

\begin{figure*}[t!]
    \centering
    \includegraphics[width=0.56\linewidth]{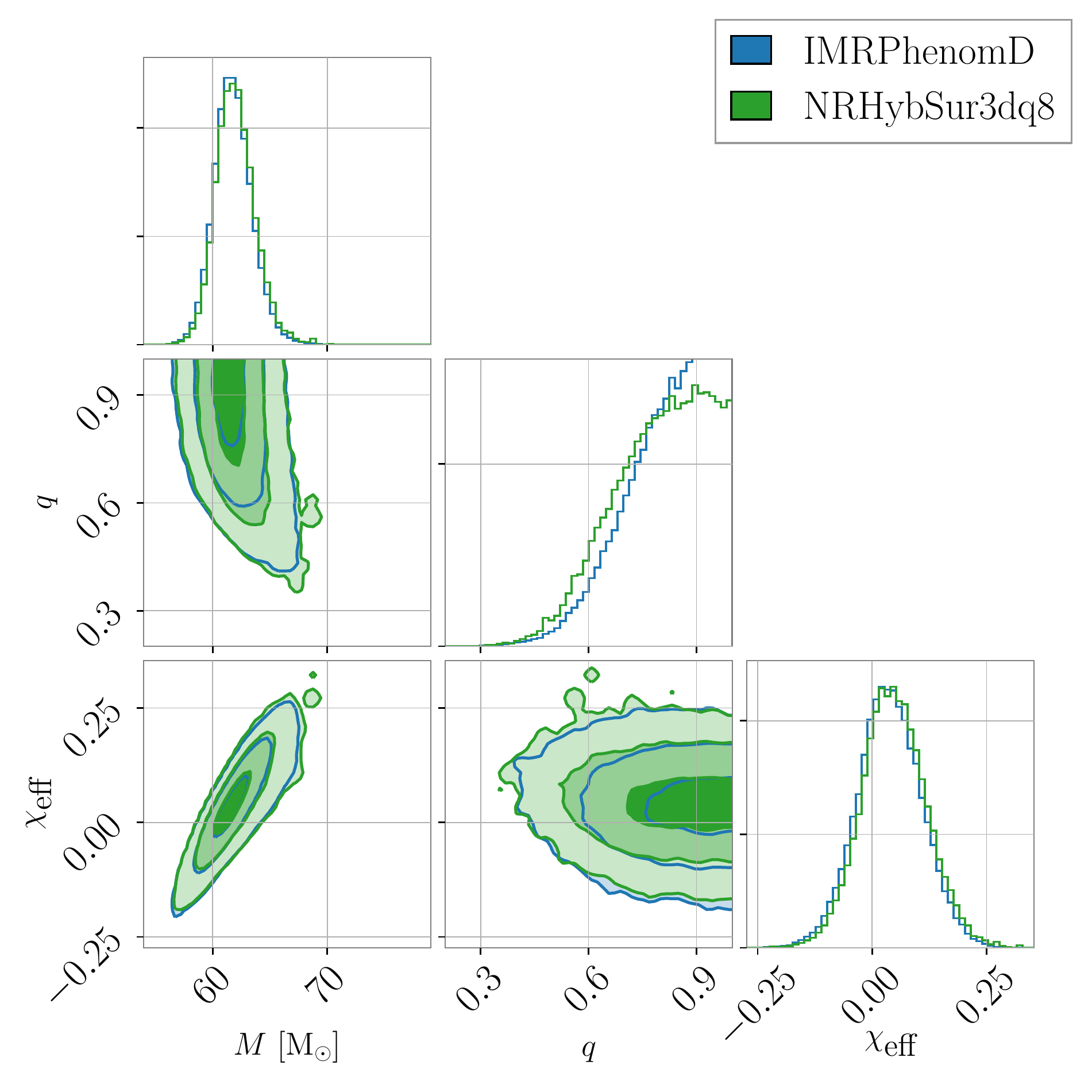}
    \includegraphics[width=0.405\linewidth]{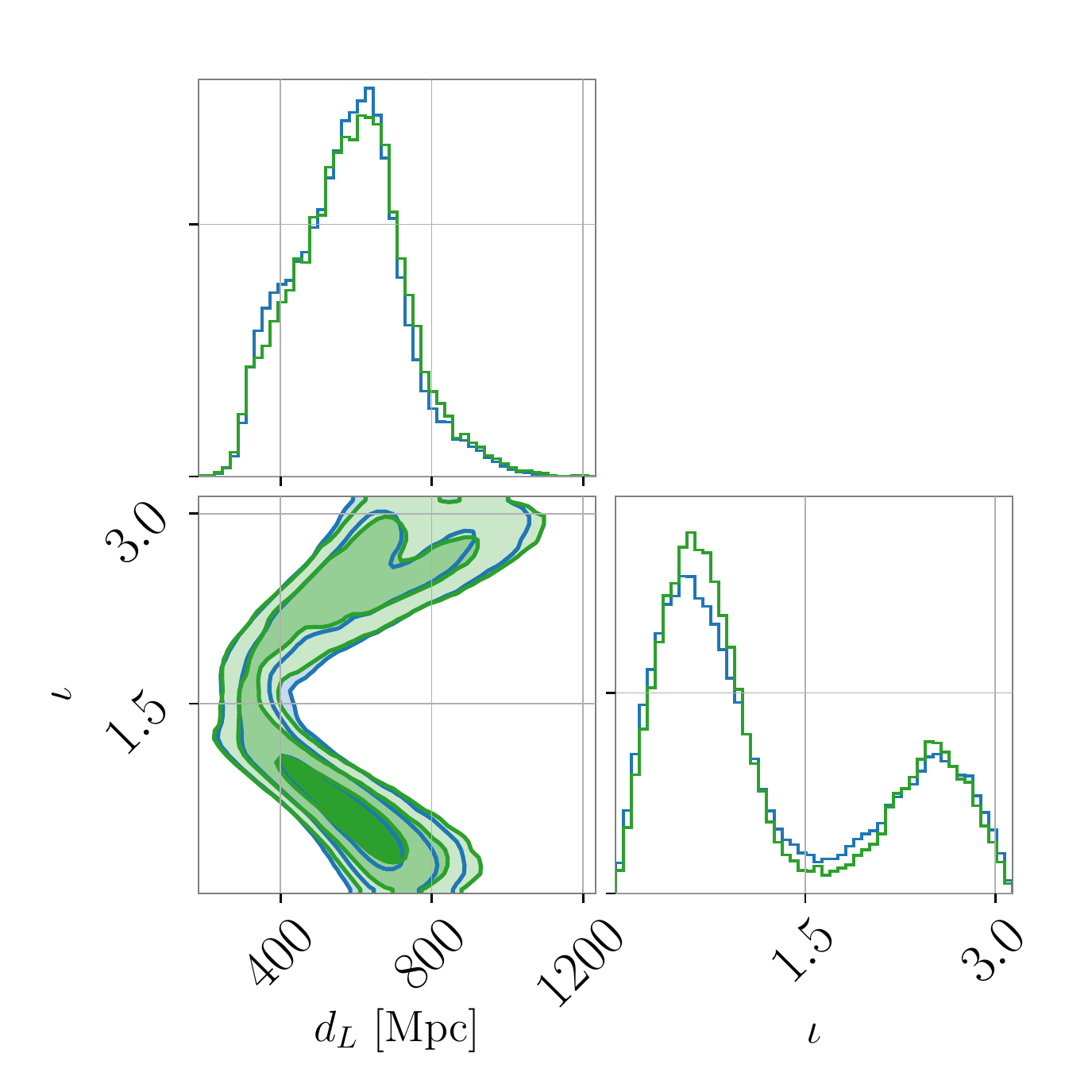}
    \caption{
    GW170814}
    \label{fig:GW170814}
\end{figure*}

\begin{figure*}[t!]
    \centering
    \includegraphics[width=0.56\linewidth]{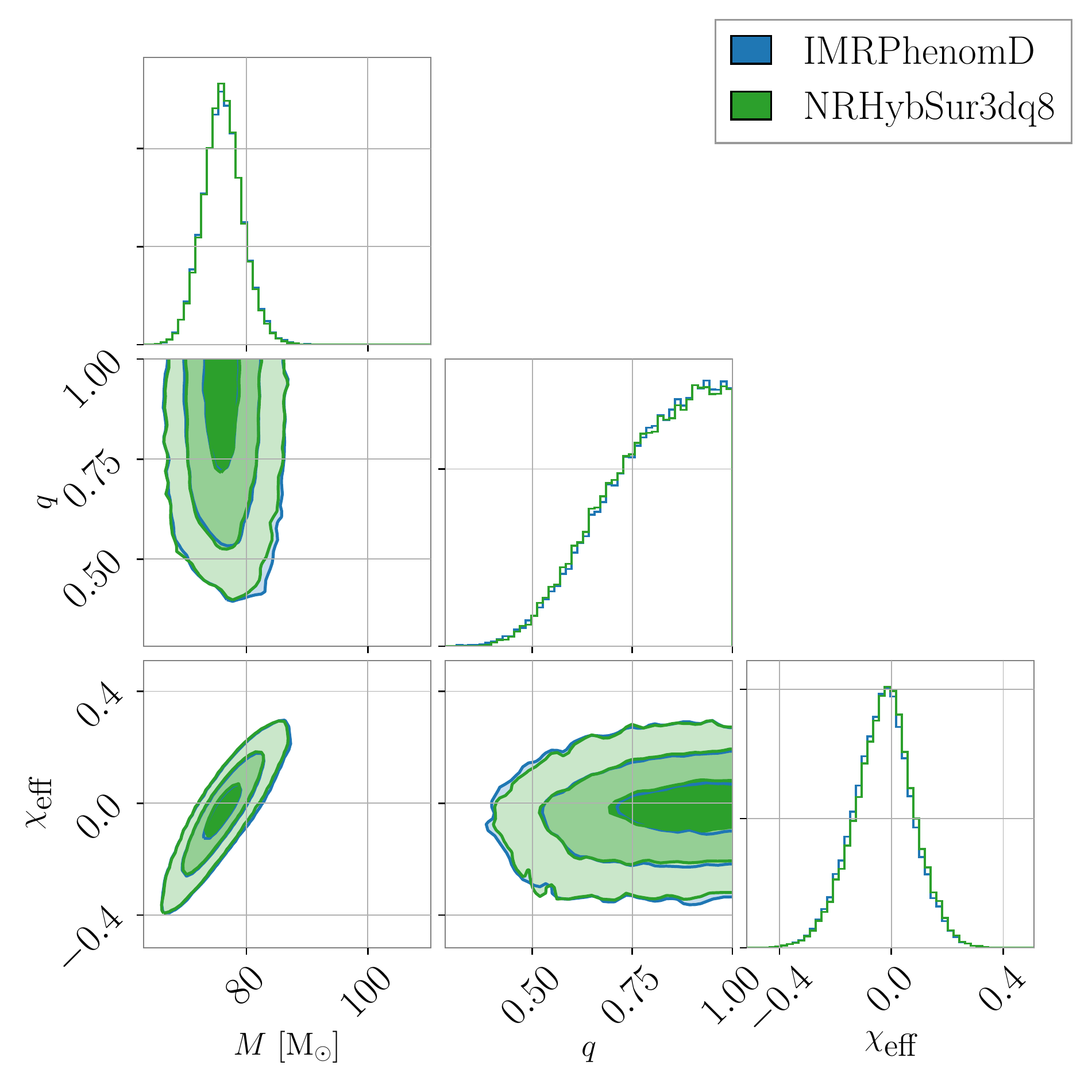}
    \includegraphics[width=0.405\linewidth]{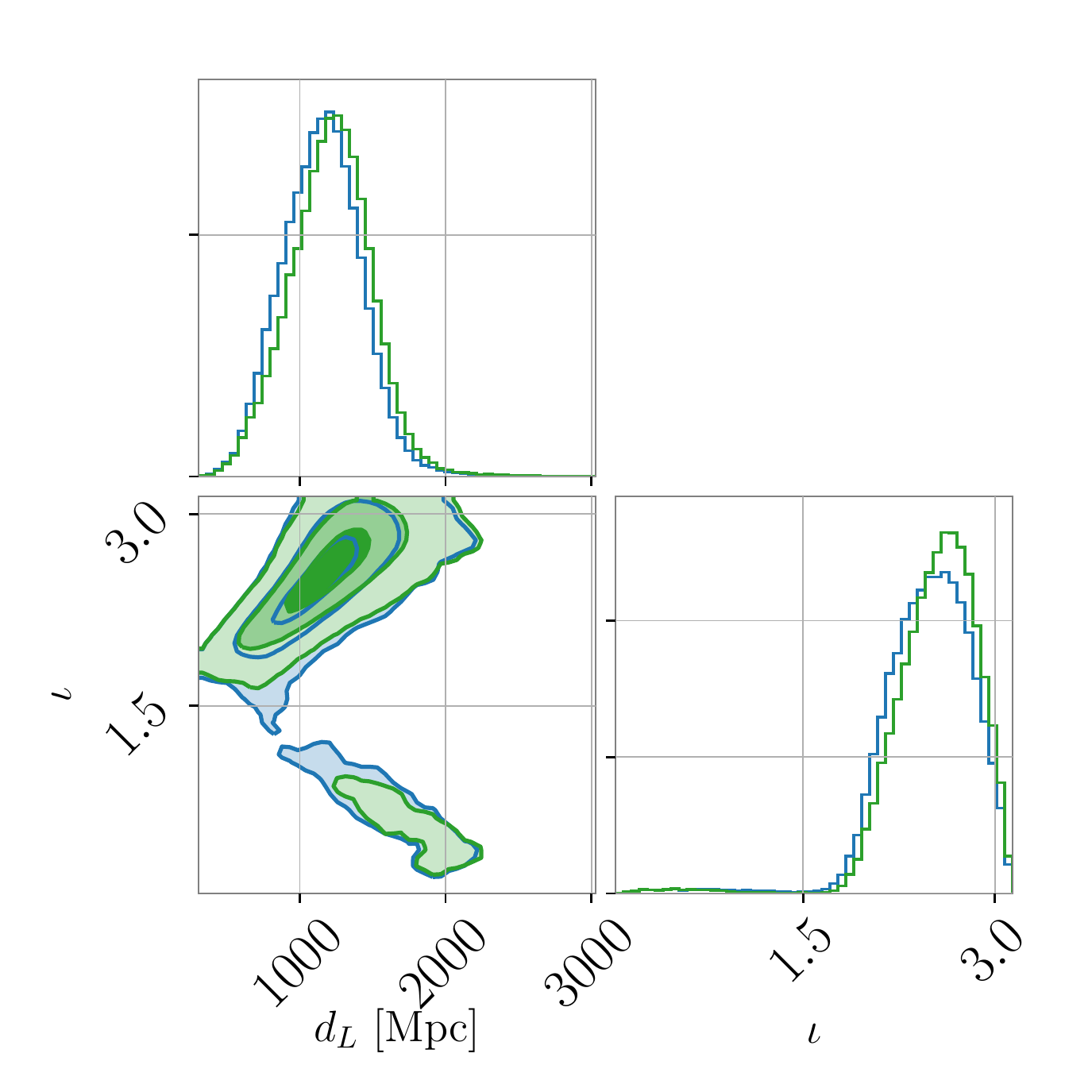}
    \caption{
    GW170818}
    \label{fig:GW170818}
\end{figure*}

\begin{figure*}[t!]
    \centering
    \includegraphics[width=0.56\linewidth]{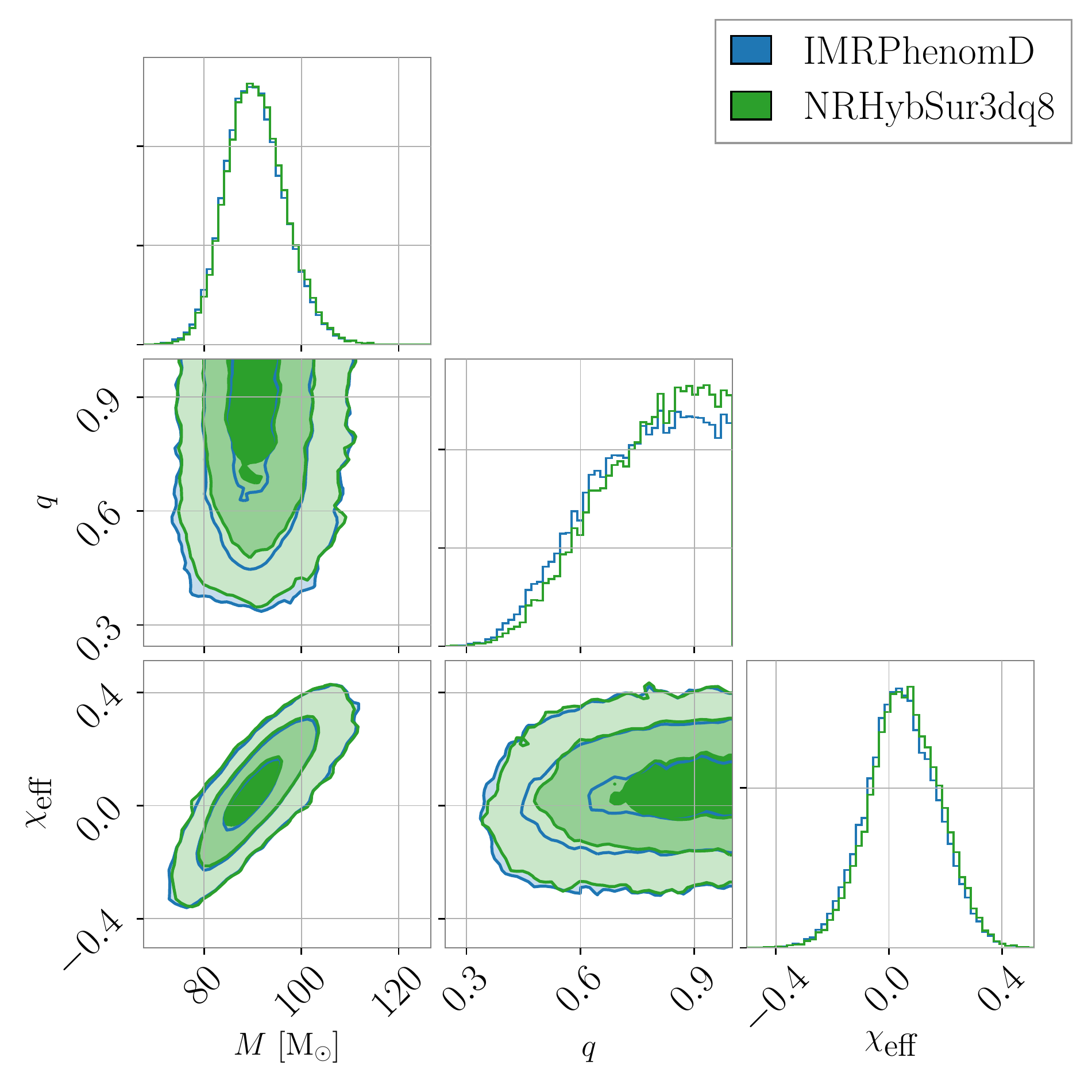}
    \includegraphics[width=0.405\linewidth]{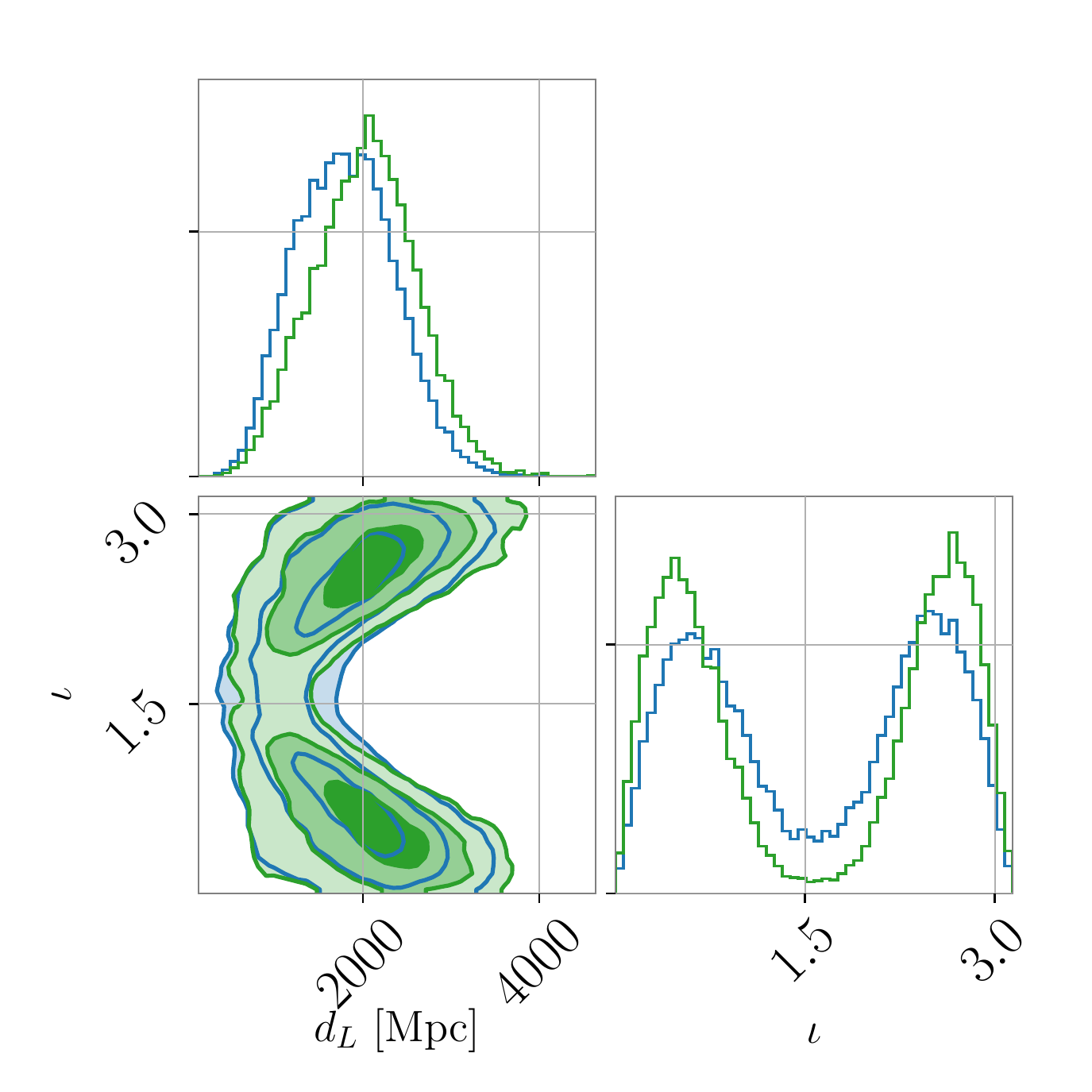}
    \caption{
    GW170823.
    This event is an example of when a negative $\ln\text{BF}$ provides information about the source.
    In this instance, the source is better constrained to be face off and therefore further away.
    This is a statement that the binary did not emit with significant higher-order modes, which are stronger for edge-on systems.
    }
    \label{fig:GW170823}
\end{figure*}

\end{appendix}

\end{document}